\documentclass[11pt]{article}

\newcommand{\bq}{\begin{quotation}\noindent}
\newcommand{\eq}{\end{quotation}}
\newcommand{\be}{\begin{equation}}
\newcommand{\ee}{\end{equation}}
\newcommand{\bea}{\begin{eqnarray}}
\newcommand{\eea}{\end{eqnarray}}
\newcommand{\bc}{\begin{center}}
\newcommand{\ec}{\end{center}}
\def\tr{{\rm tr}\,}
\def\drangle{\rangle\!\rangle}
\def\dlangle{\langle\!\langle}

\newcommand{\boxedeqn}[1]{%
  \[\fbox{%
      \addtolength{\linewidth}{-2\fboxsep}%
      \addtolength{\linewidth}{-2\fboxrule}%
      \begin{minipage}{\linewidth}%
      $$#1$$%
      \end{minipage}%
    }\]%
}

\newcommand{\veec}[3]{\left(%
\begin{array}{c}{\!\!#1\!\!}\\{\!\!#2\!\!}\\{\!\!#3\!\!}\end{array}\right)}

\newtheorem{assump}{Assumption}
\newtheorem{resump}{Resumption}

\usepackage{graphicx}
\usepackage{indentfirst}
\usepackage{amssymb}

\begin{document}

\title{Quantum-Bayesian Coherence\footnote{This is a much expanded version of a paper originally appearing as C.~A. Fuchs and R.~Schack, ``From Quantum Interference to Bayesian Coherence
and Back Round Again,'' in {\sl Foundations of Probability and Physics -- 5, V\"axj\"o, Sweden, 24--27 August 2008}, edited by L.~Accardi, G.~Adenier, C.~A. Fuchs, G.~Jaeger, A.~Khrennikov, J.~{\AA}. Larsson, and S.~Stenholm, AIP Conference Proceedings Vol.~1101, (American Institute of Physics, Melville, NY, 2009), pp.~260--279. Per always, we thank Andrei Khrennikov for organizing this wonderful series of meetings, which has been so instrumental in crystalizing our quantum-Bayesian thoughts.}}

\author{Christopher A. Fuchs$^\dagger$ and R\"udiger Schack$^\sharp$
\medskip
\\
\small
$^\dagger$Perimeter Institute for Theoretical Physics
\\
\small
%31 Caroline St.\ North,
Waterloo, Ontario N2L 2Y5, Canada
\medskip\\
\small
$^\sharp$Dept.\ of Mathematics, Royal Holloway, University of London
\\
\small
Egham, Surrey TW20 0EX, United Kingdom}

\date{11 June 2009}

\maketitle

\begin{abstract}
In a quantum-Bayesian take on quantum mechanics, the Born Rule cannot be interpreted as a rule for setting measurement-outcome probabilities from an {\it objective\/} quantum state.  But if not, what is the role of the rule?  In this paper, we argue that it should be seen as an  empirical addition to Bayesian reasoning itself.  Particularly, we show how to view the Born Rule as a normative rule in addition to usual Dutch-book coherence.  It is a rule that takes into account how one should assign probabilities to the consequences of various intended measurements on a physical system, but explicitly in terms of prior probabilities for and conditional probabilities consequent upon the imagined outcomes of a special {\it counterfactual\/} reference measurement. This interpretation is seen particularly clearly by representing quantum states in terms of probabilities for the outcomes of a fixed, fiducial symmetric informationally complete (SIC) measurement.  We further explore the extent to which the general form of the new normative rule implies the full state-space structure of quantum mechanics.  It seems to get quite far.
\end{abstract}

\tableofcontents

\section{Introduction: Unperformed Measurements Have No Outcomes}

\begin{flushright}
\baselineskip=13pt
\parbox{3.2in}{\baselineskip=13pt\small
We choose to examine a phenomenon which is impossible, {\it absolutely\/} impossible, to explain in any classical way, and which has in it the heart of quantum mechanics.  In reality, it contains the {\it only\/} mystery.  We cannot make the mystery go away by ``explaining'' how it works.  We will just {\it tell\/} you how it works.  In telling you how it works we will have told you about the basic peculiarities of all quantum mechanics.}
\medskip\\
\small --- R. P. Feynman, 1964
\end{flushright}

Richard Feynman wrote these words for the opening chapter on quantum mechanics in his monumental {\sl Feynman Lectures on Physics} \cite{Feynman64}.  It was his lead-in for a discussion of the double-slit experiment with {\it individual\/} electrons.
Imagine if you will, however, someone well-versed in the quantum foundations discussions of the last 25 years (since the Aspect experiment \cite{Aspect82}, say) yet surprisingly unaware of when Feynman wrote this.  What might he conclude Feynman was speaking of?  Would it be the double-slit experiment?  Probably not.  To the modern sensibility, a good guess would be that he was speaking of something to do with quantum entanglement or Bell-inequality violations.  In the history of foundational thinking, the double-slit experiment has fallen by the wayside.

So, what is it that quantum entanglement teaches us---via EPR-criterion-of-reality considerations and Bell-inequality violations---that the double-slit experiment does not?  A common, if quick and dirty, answer is that ``local realism'' fails.\footnote{Too quick and dirty, some would say \cite{Norsen06}. However, the conclusion drawn there---that a Bell inequality violation implies the failure of locality, full stop---is based (in part) on taking the EPR criterion of reality as sacrosanct. As will become clear in the development, it will be seen that we do not take it so.}  Unpacking the term local realism, one means more precisely the conjunction of two statements \cite{Braunstein90}:  1) that actions or experiments in one region of spacetime cannot instantaneously affect matters of fact at far away regions of spacetime, and 2) that measured values {\it pre-exist\/} the act of measurement, which merely ``reads off'' the values, rather than enacting or creating them by the process itself.  The failure of local realism means that one or the other or some combination of both these statements fails.  This, many would say, is the ``only mystery'' of quantum mechanics, or at least the only truly deep one.

But the mystery, as emphasized, has at least two sides.  It seems the majority of physicists who care about these matters think it is locality (condition 1 above) that has to be abandoned with a Bell-inequality violation---i.e., they think that there really are ``spooky actions at a distance.''\footnote{Indeed, it flavors almost everything they think of quantum mechanics, including the {\it interpretation\/} of the imaginary games or toy models they use to better understand quantum mechanics itself. Take the recent flurry of work on Popescu-Rohrlich boxes \cite{Popescu94}.  These are imaginary devices that give rise to greater-than-quantum violations of various Bell inequalities.  Importantly, another common name for these devices is the term ``nonlocal boxes'' \cite{Barrett05}.  Their exact definition comes via the magnitude of a Bell-inequality violation---which entails the non-pre-existence of values or a violation of locality or both---but the commonly used name opts only to recognize nonlocality.  They're not called anti-realism boxes, for instance.  The nomenclature is psychologically telling.}  But there is a small minority that thinks the abandonment of condition 2, that measured values pre-exist the act of measurement, is the more warranted conclusion \cite{Peres78,Wheeler82,Zeilinger96,Zukowski05,Plotnitsky06,D'Ariano08,%
Demopoulos08} and among these are the {\it quantum Bayesians\/} \cite{Fuchs01,Caves02,Schack01,Fuchs02,Fuchs03,Schack04,Fuchs04,Caves07,%
Appleby05a,Appleby05b}.\footnote{For a selection of misunderstandings on the position, see Refs.\ \cite{PhysicsTodayCriticisms,Hagar03,Grangier03,Dennis04,Marchildon04,%
Ferrero04,Hagar05,Hagar06,Mohrhoff07,Wallace07,Palge08}. \label{Misunderstandings}}$^,$\footnote{For alternative developments of several Bayesian-inspired ideas in quantum mechanics, see Refs.\ \cite{Youssef01,Pitowsky03,Pitowsky05,Srednicki05,Caticha06,Leifer06a,%
Leifer06b,Mana07,Rau07,Goyal08,Warmuth09}. We leave these citations separate because of various distinctions within each from the point of view presented here---these distinctions range from 1) the particular strains of Bayesianism each adopts, to 2) whether quantum mechanics is a {\it generalized\/} probability theory or rather simply an application within Bayesian probability per se, to 3) the level of the agent's involvement in bringing about the outcomes of quantum measurement. Nonetheless there are sometimes striking kinships between the ideas of these papers and the effort here.}  In a slogan inspired by Asher Peres's more famous one \cite{Peres78}, ``unperformed measurements have no outcomes.''

Among the several paths of argumentation the quantum Bayesians use to come to this conclusion, not least in importance is a thoroughgoing personalist account of {\it all\/} probabilities \cite{Ramsey26,DeFinetti31,Savage54,DeFinetti90,Bernardo94,Jeffrey04}---where the ``all'' in this sentence includes probabilities for quantum measurement outcomes and even the probability-1 assignments among these \cite{Caves07}.  From the quantum-Bayesian point of view, this is the only sound interpretation of probability.  Moreover, this move for quantum probabilities frees up the quantum state from any objectivist obligations.  In so doing it wipes out the mystery \cite{Einstein51,Fuchs00} of quantum-state-change at a distance \cite{Timpson08} and much of the mystery of general wave function collapse as well \cite{Fuchs02,Struggles}.

But what does all this have to do with Feynman?  Apparently Feynman too saw something of a truth in the idea that ``unperformed measurements have no outcomes.''  Yet, he did so because of considerations with regard to the double-slit experiment.  Later in the lecture he says,
\begin{quotation}
Is it true, or is it {\it not\/} true that the electron either goes through hole 1 or it goes through hole 2?  The only answer that can be given is that we have found from experiment that there is a certain special way that we have to think in order that we do not get into inconsistencies.  What we must say (to avoid making wrong predictions) is the following.  If one looks at the holes or, more accurately, if one has a piece of apparatus which is capable of determining whether the electrons go through hole 1 or hole 2, then one {\it can\/} say that it goes either through hole 1 or hole 2.  {\it But}, when one does {\it not\/} try to tell which way the electron goes, when there is nothing in the experiment to disturb the electrons, then one may {\it not\/} say that an electron goes either through hole 1 or hole 2.  If one does say that, and starts to make any deductions from the statement, he will make errors in the analysis.  This is the logical tightrope on which we must walk if we wish to describe nature successfully.
\end{quotation}
Returning to the original quote, we are left with the feeling that this is the very thing Feynman saw to be the ``basic peculiarity of all quantum mechanics.''

One should ask though, is his conclusion really compelled by so simple a phenomenon as the double slit?  How could simple ``interference'' be so far-reaching in its metaphysical implication?  Water waves interfere and there is no great mystery there.  Most importantly, the double-slit experiment is a story of measurement on a single quantum system, whereas the story of EPR and Bell is that of measurement on {\it two\/} seemingly disconnected systems.

Two systems are introduced for a good reason.  Without the guarantee of arbitrarily distant parts within the experiment---so that one can conceive of measurements on one, and draw inferences about the other---what justification would one have to think that changing the conditions of the experiment (from one slit closed to both slits open) should {\it not\/} make a deep conceptual difference to its analysis? Without such a guarantee for compelling the belief that some matter of fact stays constant in the consideration of two experiments, one---it might seem---would be quite justified in responding, ``Of course, you change an experiment, and you get a different probability distribution arising from it.  So what?$\,$\footnote{This is a point Koopman \cite{Koopman57} and Ballentine \cite{Ballentine86} seem happy to stop the discussion with.  For instance, Ballentine writes, ``One is well advised to beware of probability statements of the form, $P(X)$, instead of $P(X|C)$.  The second argument may be safely omitted only if the conditional event or information is clear from the context, and only if it is constant throughout the problem.  This is not the case in the double slit experiment. \ldots\ We observe from experiment that $P(X|C_3)\ne P(X|C_1)+P(X|C_2)$.  This fact, however, has no bearing on the validity of \ldots\ probability theory.'' \label{Gobbledygook}}''  For quite a long time, the authors thought that Feynman's logical path from example to conclusion---a conclusion that we indeed agree with---was simply misguided.  The argument just does not seem to hold to the same stringent standards as Bell-inequality analyses.

However, we have recently started to appreciate that there may be something of substance in Feynman's argument nonetheless.  It is just not something so easily seen without the proper mindset.  The key point is that the so-called ``interference'' in the example is not in a material field---of course it was never purported to be---but in something so ethereal as probability itself (a logical, not a physical, construct).  Most particularly, Feynman makes use of a beautiful and novel move:  He analyzes the probabilities in an experiment that {\it will be\/} done in terms of the probabilities from experiments that {\it won't be\/} done.  He does not simply conditionalize the probabilities to the two situations and let it go at that.\footnote{See Footnote \ref{Gobbledygook} at least once again.}  Rather he tries to see the probabilities in the two situations as functions of each other. Not functions of a condition, but functions (or at least relations) of each other.  This is an important point.   The world need not have to give a relation between these probabilities, yet it does:  Quantum mechanics is what makes the link precise.  Feynman seems to have a grasp on the idea that the essence of the quantum mechanical formalism is to provide a tool for analyzing the factual in terms of a counterfactual.\footnote{In his own case, he then ``retreats'' to the formalism of amplitudes to mediate between the various probabilities, whereas in this paper we will doggedly stick to probabilities, and only probabilities.  However the essential point is the same as the one we want to develop.  Feynman's particular formalism is thus of no matter for the present concerns; it is only the spirit that we are observing here.}

Here is the way Feynman put it in a 1951 paper titled, ``The Concept of Probability in Quantum Mechanics,'' \cite{Feynman51}:
\begin{quotation}
I should say, that in spite of the implication of the title of this talk the concept of probability is not altered in quantum mechanics.  When I say the probability of a certain outcome of an experiment is $p$, I mean the conventional thing, that is, if the experiment is repeated many times one expects that the fraction of those which give the outcome in question is roughly $p$.  I will not be at all concerned with analyzing or defining this concept in more detail, for no departure from the concept used in classical statistics is required.

What is changed, and changed radically, is the method of calculating probabilities.
%The effect of this change is greatest when dealing with objects of %atomic dimensions.
\end{quotation}
Far be it from us to completely agree with everything in this quote.  For instance, the concept of ``probability {\it as\/} long-run frequency'' is anathema to a Bayesian of any variety \cite{Good83}.  And B.~O. Koopman \cite{Koopman57} is surely right when he says,
\bq
\indent Ever since the advent of modern quantum mechanics in the late 1920's, the idea has been prevalent that the classical laws of probability cease, in some sense, to be valid in the new theory. More or less explicit statements to this effect have been made in large number and by many of the most eminent workers in the new physics.\footnote{In fact, Koopman is speaking directly of Feynman here.  Moreover, both he and Ballentine \cite{Ballentine86} have criticized Feynman on the same point: That with his choice of the word ``changed'' in the last quote, Feynman implicates himself in not recognizing that the conditions of the three contemplated experiments are distinct and, hence, in not conditionalizing his probabilities appropriately.  Thus---Koopman and Ballentine say---it is no wonder Feynman thinks he sees a violation of the laws of probability.

In the authors' opinion however, Koopman and Ballentine are hanging too much on the word ``changed''---we rather see it as an unfortunate choice of wording on Feynman's part.  That he understood that the conditions are different in a deep and inescapable way in the three contemplated experiments, we feel, is documented well enough in the quote above from his 1964 lecture.} \ldots\

Such a thesis is surprising, to say the least, to anyone holding more or less conventional views regarding the positions of logic, probability, and experimental science: many of us have been apt---perhaps too naively---to assume that experiments can lead to conclusions only when worked up by means of logic and probability, whose laws seem to be on a different level from those of physical science.
\eq
But there is a kernel of truth here that should not be dismissed in spite of Feynman's diversion to frequentism and his poor choice of the word ``changed'':
\begin{center}
\parbox{5.2in}{\it The concept of probability is not altered in quantum mechanics (it is personalistic Bayes\-ian probability
like always).  What is radical is the recipe it gives for calculating new probabilities from old.}
\end{center}

For, quantum mechanics---we plan to show in this paper---gives a resource that raw Bayesian probability theory does not:  It gives a rule for forming probabilities for the outcomes of {\it factualizable\,}\footnote{We coin this term because it stands as a better counterpoint to the term ``counterfactual'' than the term ``actualizable'' seems to.  We also wanted to capture the following idea a little more plainly:  Both measurements being spoken of here are only potential measurements---it is just that one will always be considered in the imaginary realm, whereas the other may one day become a fact of the matter if it is actually performed.} experiments (experiments that may actually be performed) from the probabilities one assigns for the outcomes of a designated {\it counterfactual\/} experiment (an experiment only imagined, and though possible to do, never actually performed).  So, yes, unperformed measurements have no outcomes as Peres has expressed nicely; nonetheless, imagining their performance can aid in analyzing the probabilities one {\it ought to\/} assign for an experiment that may be factually performed.  Putting it more carefully than Feynman:  Quantum mechanics does not provide a radical change to the method of calculating probabilities; it provides rather an empirical {\it addition\/} to the laws of Bayesian probability.

In this paper, we offer a modernization and Bayesianification of the Feynman point by making intimate use of a beautiful representation of quantum states that was not available in his time\footnote{And is not {\it yet\/} really available at the present either.  But this will be explained in Section 3.}:  It is one based on SICs or symmetric informationally complete observables \cite{Caves99,Zauner99,Renes04,Fuchs04b,Appleby05,ApplebyDangFuchs}.  The goal is to make it more transparent than ever that the very content of the Born Rule is not that it gives a procedure for {\it setting\/} probabilities (from some independent entity called ``the quantum state''), but that it represents a surprising ``method of calculating probabilities,'' new ones from old.

That this must be the meaning of the Born Rule more generally in the quantum-Bayesian approach to quantum foundations has been argued from several angles by the authors before \cite{Caves07,Struggles}.  For instance, in Ref.~\cite{Caves07}, we put the point this way:
\bq
\indent We have emphasized that one of the arguments often repeated to justify that quan\-tum-mechanical probabilities are objective, rather than subjective, is that they are ``determined by physical law.''  But, what can this mean?  Inevitably, what is being invoked is an idea that quantum states $|\psi\rangle$ have an existence independent of the probabilities they give rise to through the Born rule, $p(i)=\langle\psi| E_i | \psi\rangle$. From the Bayesian perspective, however, these expressions are not independent at all, and what we have argued in this paper is that quantum states are every bit as subjective as any Bayesian probability.  What then is the role of the Born rule?  Can it be dispensed with completely?

It seems no, it cannot be dispensed with, even from the Bayesian
perspective.  But its significance is different than in other developments of quantum foundations: The Born rule is not a rule for {\it setting\/} probabilities, but rather a rule for {\it transforming\/} or {\it relating\/} them.

For instance, take a complete set of $d+1$ observables $O^k$,
$k=1,\ldots,d+1$, for a Hilbert space of dimension $d$.  Subjectively setting probabilities for the $d$ outcomes of each such measurement uniquely determines a quantum state $|\psi\rangle$ (via inverting the Born rule).  Thus, as concerns probabilities for the outcomes of any other quantum measurements, there can be no more freedom.  All further probabilities are obtained through linear transformations of the originals.  In this way, the role of the Born rule can be seen as having something of the flavor of Dutch-book coherence, but with an empirical content added on top of bare, law-of-thought probability theory: An agent interacting with the quantum world would be wise to adjust his probabilities for the outcomes of various measurements to those of quantum form if he wants to avoid devastating consequences. The role of physical law---i.e., the assumption that the world is made of quantum mechanical stuff---is codified in how measurement probabilities are related, not how they are set.
\eq
What is new in the present paper is the emphasis on a single designated observable for the counterfactual thinking, as well as a detailed exploration of the rule for combining probabilities in this picture.  Particularly, we will see that a significant part of the structure of quantum-state space arises from the consistency of that rule---a single formula we designate {\it the urgleichung}.  The urglei\-chung is the stand-in (and correction) in our context for Feynman's flawed but nonetheless forceful sentence, ``What is changed, and changed radically, is the method of calculating probabilities.''

Returning to the point we started our discussion with, the one about interference going by the wayside in quantum foundations, we should say the following.  To the extent that the full formalism of quantum mechanics can be re-derived from a simple Feynman-style scenario---even if not the double-slit experiment per se, but nonetheless one wherein probabilities for the outcomes of factualizable experiments are obtained from probabilities in a designated counterfactual one---that scenario must indeed express the essence of quantum mechanics.  For if these considerations give rise to the full formalism of the theory (Hilbert spaces, positive operators, the possibility of tensor-product decompositions, etc.), they must give rise to entanglement, Bell-inequality violations, and Kochen-Specker `paradoxes' as well:  These will be established as corollaries to the formalism.  That is to say, hidden in the womb of these considerations would be every mystery and every `paradox' of quantum mechanics.  And if that truly is the case, who could be bold enough to say that the simple scenario does not carry in it the essence of quantum mechanics?  From our point of view, it goes without saying that the exploration of quantum mechanics' ability to engender Bell-inequality violations and Kochen-Specker theorems is an immensely instructive activity for sorting out the implications of the theory.  Nonetheless, one should not lose sight of the potential loss of understanding that can be incurred by confusing a corollary with a postulate.  In this sense, Feynman may well have had the right intuition.

\subsection{Outline of the Paper}

The plan of the paper is as follows.  In Section 2, we review the personalist Bayesian account of probability, showing how some Dutch-book arguments work, and emphasizing a point we have not seen emphasized before:  Bayes Rule and the Law of Total Probability are not necessities in a Bayesian account of probability.  Those rules are operative when there is a {\it conditional lottery\/} in the picture that can actually be gambled upon.  When there is no such lottery, the rules hold no force---indeed, there is no means to even define the terms in them.  One can introduce an imaginary conditional lottery to give the terms some substance, but then there is nothing in Dutch-book coherence itself that can be used to compel the usual equations.

In Section 3, we review the notion of a SIC and show a new sense in which it is a very special measurement. Most importantly we delineate the full structure of quantum-state space in SIC terms. From one point of view the derived formalism is somewhat ugly in comparison to the usual one of complex vector spaces and operators, but from another point of view it is surprisingly beautiful:  Particularly, by making use of a SIC instead of any other informationally complete measurement, the formalism is as simple and compact as it can be subject to these considerations.  We also make a point to show that unitary time evolution is, in fact, one aspect of quantum mechanics where no essential complication is introduced at all:  up to a very small change of formula, unitary time evolution when written in SIC terms, looks almost identical to classical stochastic evolution.

Section 4 contains the heart of the paper.  In it, we introduce the idea of thinking of an imaginary (counterfactual) SIC behind all quantum measurements, so as to give an imaginary conditional lottery with which to define conditional probabilities.  We then show how to write the Born Rule in these terms, and find it to appear strikingly similar to the Law of Total Probability.  We then show how this move in interpretation is radically different from the one offered by the followers of ``objective chance'' in the Lewisian sense. Finally, we discuss how these arguments supply a ground for taking quantum mechanics to be based on the complex number field rather the real numbers.

In Section 5, we show that one can derive at least some of the features of quantum-state space by taking this modified or {\it Quantum\/} Law of Total Probability as a postulate.  Particularly, we show that with a couple of smaller assumptions, it gives rise to a generalized Bloch sphere and defines an underlying ``dimensionality'' for a system that matches the one given by the true quantum mechanical state space. We also demonstrate a few other features of the geometry these considerations give rise too.

In Section 6, we step back still further in our considerations and explore the extent to which the particular constants $d^2$, $d+1$, and $\frac{1}{d}$ in our Quantum Law of Total Probability can arise from even more elementary considerations.  This section is a preliminary attempt to understand the origin of the equation treated simply as a postulate in Section 4.  Unfortunately, its results at this stage remain more mathematical than insightful, but they do give an indication of how the whole structure seems to hang together rather tightly.

In Section 7 we give a brief discussion of where we stand at this stage of research.  Finally in Section 8, ``Forward to an Eventual Foreword!,'' we close the paper by discussing how our work is still far from done:  Hilbert space, from a quantum-Bayesian view, has not yet been derived, only indicated.  Nonetheless the progress made here gives us hope that we are inching our way toward a formal expression of the ontology underlying a quantum-Bayesian vision of quantum mechanics:  It really does have to do with the Peres slogan, but tempered with a kind of `realism' that Peres would probably not have accepted forthrightly.\footnote{We say this because of Peres's openly acknowledged positivist tendencies.  See Chapters 22 and 23 in Ref.~\cite{Fuchs01} where he would sometimes call himself a ``recalcitrant positivist.''  Also see the opening remarks of Ref.~\cite{Peres03} for a good laugh.}  On the other hand, it is not a `realism' that we expect to be immediately accepted by most modern philosophers of science either.\footnote{See Footnote \ref{Misunderstandings} and Ref.~\cite[Sec.~4.1]{Timpson08} for a sampling of instances of opposition to our inchoate ontology, and see Refs.~\cite{Nagel89,Dennett04,Price97} for details on the ``view from nowhere''/``view from nowhen'' {\it weltanschauung\/}  more generally.}  This is because it is already clear that whatever it will ultimately turn out to be, it is based on 1) a rejection of the ontology of the block universe \cite{James1882,James1884,James1910}, and 2) a rejection of the ontology of the detached observer \cite{Pauli94,Laurikainen88,Gieser05}.  The `realism'  of vogue in philosophy of science circles, which makes heavy use of both these elements, is too narrow a concept for our purposes.  Reality, the stuff of which the world is made, the stuff that was here before Darwinian evolution brought about agents and observers, the stuff that prevents us from vanishing in a dream---it strikes us---is more interesting than that.

\section{Personalist Bayesian Probability}

From the Bayesian point of view, probability is degree of belief as measured by action.  More precisely, we say one has (explicitly or implicitly) assigned a probability $p(A)$ to an event $A$ if, before knowing the value of $A$, one is willing to either {\it buy\/} or {\it sell\/} a lottery ticket of the form
\begin{center}
\parbox{2.0in}{\boxedeqn{\mbox{Worth \$1 if $A$.}}}
\end{center}
for an amount $\$ p(A)$.  The personalist Bayesian position adds only that this is the full meaning of probability; it is nothing more and nothing less than this definition.  Particularly, nothing intrinsic to the event or proposition $A$ can help declare $p(A)$ right or wrong, or more or less rational.  The value $p(A)$ is solely a statement about the agent who holds it.

Nonetheless, even for a personalist Bayesian, probabilities do not wave in the wind.  Probabilities are held together by a normative principle: That whenever an agent declares probabilities for various events---say $A$, $B$, $A\vee B$, $A\wedge B\vee C$, etc.---he should {\it strive\/} to never gamble (i.e., buy and sell lottery tickets) so as to incur what he believes will be a sure loss.  This normative principle is known as Dutch-book coherence.  And from it, one can derive the usual calculus of probability theory.

This package of views about probability (that in value it is personal, but that in function it is akin to the laws of logic) had its origin in the mid-1920s and early 1930s with the work of F.~P. Ramsey \cite{Ramsey26} and B. de Finetti \cite{DeFinetti31}.  J. M. Keynes put both crucial aspects of the view very succinctly \cite{Keynes51}:
\begin{quotation}
\noindent The application of these ideas [regarding formal logic] to the logic of probability is very fruitful.  Ramsey argues, as against the view which I had put forward, that probability is concerned not with objective relations between propositions but (in some sense) with
degrees of belief, and he succeeds in showing that the calculus of
probabilities simply amounts to a set of rules for ensuring that the
system of degrees of belief which we hold shall be a {\it
consistent\/} system.  Thus the calculus of probabilities belongs to
formal logic.  But the basis of our degrees of belief---or the {\it a
priori}, as they used to be called---is part of our human outfit,
perhaps given us merely by natural selection, analogous to our
perceptions and our memories rather than to formal logic.
\end{quotation}
And this quote of B.~O. Koopman refines the first aspect very nicely \cite{Koopman40}:
\bq
\indent The intuitive thesis in probability holds that both in its meanings and in the laws which it obeys, probability derives directly from the intuition, and is prior to objective experience; it holds that it is for experience to be interpreted in terms of probability and not for probability to be interpreted in terms of experience \ldots
%``(Objective) experience'' is used here practically as the equivalent %of ``laboratory experiment'' in the narrow objective sense of the %word, and excludes introspective or ``subjective experience.'' %Moreover, it is a question here of {\it rational\/} derivation from %experience, and the {\it a priori\/} view of probability is quite %consistent with the idea that probability as well as logic may be %derived by race experience through the process of evolution.
\eq

Let us go through some of the derivation of the probability calculus from Dutch-book coherence so that we may better make a point concerning quantum mechanics afterward.\footnote{Here we basically follow the development in Richard Jeffrey's posthumously published book {\sl Subjective Probability, The Real Thing} \cite{Jeffrey04,Skyrms87}, but with our own emphasis.}  First we establish that our normative principle requires $0\le P(A)\le 1$.  For suppose $P(A)<0$.  This means an agent will sell a ticket for negative money---i.e., he will pay someone $\$ p(A)$ to take the ticket off his hands.  Regardless of whether $A$ occurs or not, the agent will then be sure he will lose money.  This violates the normative principle.  Now, take the case $P(A)>1$.  This means the agent will buy a ticket for more than it is worth even in the best case---again a sure loss for him and a violation of the normative principle.  So, probability in the sense of ticket pricing should obey the usual range of values.

Now let us establish the probability sum rule.  Suppose our agent believes two events $A$ and $B$ to be mutually exclusive---i.e., he is sure that if $A$ occurs, $B$ will not, or if $B$ occurs, $A$ will not. We can contemplate three distinct lottery tickets:
\begin{center}
\parbox{2.0in}{\boxedeqn{\mbox{Worth \$1 if $A\vee B$.}}}
\end{center}
and
\begin{center}
\parbox{2.0in}{\boxedeqn{\mbox{Worth \$1 if $A$.}}}
%\end{center}
\qquad\quad
%\begin{center}
\parbox{2.0in}{\boxedeqn{\mbox{Worth \$1 if $B$.}}}
\end{center}
Clearly the value of the first ticket should be same as the total value of the other two.  For instance, suppose an agent had set $P(A\vee B)$, $P(A)$ and $P(B)$ such that $P(A\vee B) > P(A) + P(B)$.  Then---by definition---when confronted with a seller of the first ticket, he must be willing to buy it, and when confronted with a buyer of the other two tickets, he must be willing to sell them.  But then the agent's initial balance sheet would be negative: $-\$ P(A\vee B) + \$P(A) + \$P(B)<\$0$.  And whether $A$ or $B$ or neither event occurs, it would not improve his finances: If a dollar flows in (because of the bought ticket), it will also flow out (because of the agent's responsibilities for the sold tickets), and still the balance sheet is negative.  The agent is sure of a loss.  A similar argument goes through if the agent had set his ticket prices so that $P(A\vee B) < P(A) + P(B)$.  Thus whatever values are set, the normative principle prescribes that it had better be the case that $P(A\vee B) = P(A) + P(B)$.

Consider now the following lottery ticket of a slightly different structure:
\begin{center}
\parbox{2.0in}{\boxedeqn{\mbox{Worth \,{\$}$\frac{m}{n}$\, if $A$.}}}
\end{center}
where $m\le n$ are integers. Does Dutch-book coherence say anything about the value of this ticket in comparison to the value of the standard ticket---i.e., one worth \$1 if $A$?  It does.  An argument quite like  the one above dictates that it should be valued {\$}$\frac{m}{n}P(A)$.  If a real number $\alpha$ were in place of the $\frac{m}{n}$ as similar result follows from a limiting argument.

Now we come to the most interesting and important case.  Bayesian probability is not called by its name for lack of a good reason.  A key rule in probability theory is Bayes' rule relating joint to conditional probabilities:
\be
p(A\wedge B)=p(A)p(B|A)\;.
\label{BayesBoy}
\ee
Like the rest of the structure of probability theory within the Bayesian conception, this rule must arise from an application of Dutch-book coherence.  What is that application?

The only way anyone has seen how to do it is to introduce the idea of a {\it conditional lottery}.  In such a lottery, the value of the event $A$ is revealed to the agent first.  If $A$ obtains, the lottery proceeds to the revelation of event $B$, and finally all monies are settled up.  If on the other hand $\neg A$ obtains, the remainder of the lottery is called off, and the monies put down for any ``conditional tickets'' are returned.  That is to say, the meaning of $p(B|A)$ is taken to be the price {\$}$p(B|A)$ at which one is willing to {\it buy\/} or {\it sell\/} a lottery ticket of the following form:

\begin{center}
\parbox{3.5in}{\boxedeqn{\mbox{Worth \,{\$}1 if $A\wedge B$.
\quad But return price if $\neg A$.
}}}
%\medskip\\
\qquad\quad for price {\$}$p(B|A)\;$.
\end{center}
Explicitly inserting the definition of $p(B|A)$, this becomes:
\begin{center}
\parbox{3.5in}{\boxedeqn{\mbox{Worth \,{\$}1 if $A\wedge B$.
\quad But return {\$}$p(B|A)$ if $\neg A$.
}}}
\end{center}
Now comes the coherence argument.  For, if you think about it, the price for this ticket had better be the same as the total price for these two tickets:
\begin{center}
\parbox{2.0in}{\boxedeqn{\mbox{Worth \$1 if $A\wedge B$.}}}
\qquad\quad
\parbox{2.0in}{\boxedeqn{\mbox{Worth {\$}$p(B|A)$ if $\neg A$.}}}
\end{center}
That is to say, to guard against a sure loss, we must have
\bea
p(B|A) &=& p(A\wedge B) + p(B|A)p(\neg A) \nonumber
\\
&=& p(A\wedge B) + p(B|A) - p(B|A) p(A)\;.
\eea
Consequently, Eq.~(\ref{BayesBoy}) should hold whenever there is a conditional lottery under consideration.

\subsection{When a Conditional Lottery Is Not Without Consequence}

\label{OhBoy}

But what if no conditional lottery is on the table?  Is there still any sense to conditional probabilities and Bayes' rule?  None whatsoever, and this is an important point.  Without a conditional lottery, there is no good meaning for the symbol $p(B|A)$.

Let us ask a more subtle variation on the question, then.  Suppose there is some phenomenon of importance to an agent that takes on values $B$, and he judges that before the event $B$ is revealed, a different event with mutually exclusive values $A$ will be revealed.  A Dutch bookie asks him to commit on various unconditional and conditional lottery tickets.  What can we say of the probabilities he ought to ascribe?  A minor variation of the Dutch-book arguments above tells us that whatever $p(A)$'s, $p(B)$'s, and $p(B|A)$'s he commits to, they ought---if he is is coherent---satisfy the Law of Total Probability:
\be
p(B)=\sum_A p(A) p(B|A)\;.
\label{TotalProb}
\ee
Imagine then that he does indeed commit to values satisfying this relation.

Here's the subtler question:  What now if the rug is pulled from underneath him, and it is revealed that there will be no event $A$ after all?  There will only be the event $B$. Is the agent still normatively committed to buying and selling $B$-lottery tickets for the price {\$}$p(B)$ in Eq.~(\ref{TotalProb}) that he originally expressed?  Not at all!  That would clearly be silly in some situations, and no clear-headed Bayesian would ever imagine otherwise.  The bringing about of event $A$ might so change the situation for bringing about $B$ that he simply would not gamble on it in the same way.  To hold fast to the {\$}$p(B)$ valuation of a $B$-lottery ticket, then, is not a necessity enforced by coherence, but a judgment that might or might not be the right thing to do.

In fact, one might regard the holding fast to the initial value {\$}$p(B)$ in spite of the nullification of the conditional lottery as the formal definition of precisely what it means to judge an unperformed measurement {\it to have\/} an outcome.  It means one judges that looking at the value of $A$ is incidental to the whole affair, and this is reflected in the way one gambles on $B$.  So, if $q(B)$ represents the probabilities with which the agent gambles supposing the $A$-lottery nullified, then a formal statement of the Peresian slogan that the unperformed $A$-measurement had no outcome (i.e., measuring $A$ matters, and it matters even if one makes no note of the outcome) is that
\be
q(B)\ne p(B)\;.
\ee

Still, one might imagine situations in which even if an agent judges that equality does not hold for them, he nonetheless judges that $q(B)$ and $p(B)$ should bear a precise relation to each other.  For instance, it might be the case that he judges that there is a simple function $F:[0,1]\rightarrow [0,1]$ such that
\be
q(B)=F(p(B))\;.
\label{Moufa!}
\ee
To give a somewhat concrete example, $F$ might be such that it accentuates probabilities, taking small valuations to even smaller ones and relatively large ones to ones larger still.  The only obvious consistency requirement is that it had better be the case that $F(p(B))\ge 0$ for all $B$, and $\sum_B F(p(B))=1$.  There is no necessity, however, that $F$ need take on the full interval $[0,1]$ for its domain of definition. Particularly, in a case of accentuation like the one described, it might be the case that $F$'s structure is such that it cannot be fed an arbitrary value for $p(B)$.  If that is so, one is adopting by the judgement of (\ref{Moufa!}) an indirect restriction on the values one would consider for $p(B)$, and thus indirectly one's $p(A)$ and $p(B|A)$.  That is to say, if one really, truly posits an $F$, then when gambling on the factual, one is committed to keeping an eye toward the counterfactual.

In the same vein, the agent might judge a somewhat more complicated affair that depends on the previously defined set of conditional probabilities $\{p(B|A)\}$ for the various values of $A$, like
\be
q(B)=G\Big(p(B),\{p(B|A)\}\Big)\;.
\label{Noufa!}
\ee
In Section 4, we will show that, in fact, the positive content of the Born Rule as an addition to Bayesianism is a pair of relations of these varieties, Eqs.~(\ref{Moufa!}) and (\ref{Noufa!})---the first captures the story for von Neumann measurements, the second more generally for positive-operator valued measurements.

\section{Expressing Quantum-State Space in Terms of SICs}

Let ${\mathcal H}_d$ be a finite-dimensional Hilbert space (with dimension $d$) associated with some physical system.  A quantum state for the system is usually expressed as a unit-trace positive semi-definite linear operator $\rho\in{\mathcal L}({\mathcal H}_d)\,$. \ However, we can provide an injective mapping between the convex set of these operators and the set of probability distributions\footnote{Please note our pseudo-Dirac notation $\|v\drangle$ for vectors in a {\it real\/} vector space of $d^2$ dimensions.  The relevant probability simplex for us---the one we are mapping quantum states $\rho$ to, denoted $\Delta_{d^2}$---is a convex body within this linear vector space.  Thus, its points may be expressed with the notation $\|p\drangle$ as well.  The choice of a pseudo-Dirac notation for probability distributions also emphasizes that one should think of the valid $\|p\drangle$ as a direct expression of the set of quantum states.}
\be
\| p\drangle=\Big(p(1), p(2), \ldots, p(d^2)\Big)^{\!\rm T}
\ee
over $d^2$ outcomes---the probability simplex $\Delta_{d^2}$---by first fixing a so-called (minimal) informationally complete fiducial measurement $\{E_i\}$, $i=1,\ldots,d^2$.  This is a set of positive semi-definite operators $E_i$ such that $\sum_i E_i=I$---that is, a positive-operator valued measure or POVM---but with all the $E_i$ linearly independent.  With respect to such a measurement, the probabilities $p(i)$ for its outcomes completely specify $\rho$. This follows because the $E_i$ form a basis for ${\mathcal L}({\mathcal H}_d)$, and the probabilities $p(i)=\tr \rho E_i$ can be viewed as instances of the Hilbert-Schmidt inner product $(A,B)=\tr A^\dagger B$. \ The quantities $p(i)$ thus merely express the projections of the vector $\rho$ onto the basis vectors $E_i$.  These projections completely fix the vector $\rho$.

\begin{figure} %\leavevmode
\begin{center}
\includegraphics[height=2in]{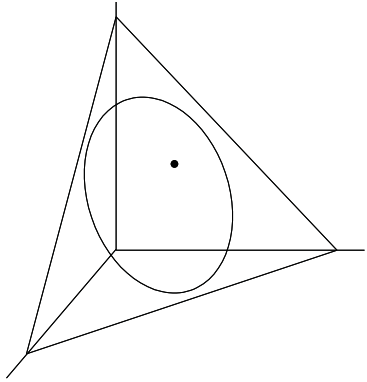}
\bigskip\caption{The planar surface represents the convex set of all
probability distributions over $d^2$ outcomes---the probability simplex $\Delta_{d^2}$.  The probability distributions valid for representing the set of quantum states, however, is a smaller convex set within the simplex---here depicted as an ellipsoid.  In the real world outside this depiction, the convex shape is quite complex.  The choice of a SIC for defining the mapping, however, makes the shape as simple as it can be with respect to the natural geometry of the simplex.}
\end{center}
\end{figure}

One can see how to calculate $\rho$ in terms of the vector $\|p\drangle$ in the following way.  Since the $E_i$ form a basis, there must be some expansion $\rho=\sum \alpha_j E_j$, where the $\alpha_j$ are real numbers making up a vector $\|\alpha\drangle$. \ Thus,
\be
p(i)=\sum_j \alpha_j\, \tr E_i E_j\;.
\ee
If we let a matrix $M$ be defined by entries $M_{ij}=\tr E_i E_j$, this just becomes
\be
\|p\drangle = M \|\alpha\drangle\;.
\ee
Using the fact that $M$ is invertible because the $E_i$ are linearly independent, we have finally
\be
\|\alpha\drangle = M^{-1} \|p\drangle\;.
\label{YakBoy}
\ee

The most important point of this exercise is that with such a mapping established, one has every right to think of a quantum state as a probability distribution {\it full stop}.  Mathematically, it is certainly nothing more, and in Ref.~\cite{Fuchs02} it is argued that conceptually it is nothing more as well.  However, it is important to note that the mapping $\rho\mapsto \|p\drangle$, though injective, cannot be surjective.  In other words, only some probability distributions in the simplex are valid for representing quantum states. A significant part of understanding quantum mechanics is understanding the range of shapes available under these mappings \cite{Bengtsson06}.

Particularly, it is important to recognize that informationally complete measurements abound---they come in all forms and sizes. Hence there is no unique representation of quantum states of this variety.  A reasonable question thus follows: What is the best measurement one can use for a mapping $\rho\mapsto \|p\drangle\,$? \ One would not want to unduly burden the representation with extra terms and calculations if one does not have to.  For instance, it would be beautiful if one could take the informationally complete measurement $\{E_i\}$ so that $M$ is simply a diagonal matrix or even the identity matrix itself.  Such an extreme simplification, however, is not in the cards---it cannot be done.

Its failure does, however, point to an interesting direction for development:  If one cannot make $M$ diagonal, one might still want to make $M$ as close to the identity as possible.  A convenient measure for how far $M$ is from the identity is the squared Frobenius distance:
\bea
F &=& \sum_{ij} \Big(\delta_{ij}-M_{ij}\Big)^2\nonumber
\\
&=& \sum_i \Big(1-\tr E_i^2\Big)^2 + \sum_{i\ne j} \Big(\tr E_i E_j\Big)^2\;.
\eea
We can place a lower bound on this quantity with the help of a special instance of the Schwarz inequality:  If
$\lambda_r$ is any set of $n$ nonnegative numbers, then
\be
\sum_r
\lambda_r^2 \ge \frac{1}{n}\!\left(\sum_r \lambda_r\right)^{\! 2}\;,
\ee
with equality if and only if $\lambda_1 = \dots = \lambda_n$.  Thus,
\be
F\ge \frac{1}{d^2}\left(\sum_i \Big(1-\tr E_i^2\Big)\right)^2 + \frac{1}{d^4-d^2}\left(\sum_{i\ne j} \tr E_i E_j\right)^2\;.
\ee
Equality holds in this if and only if there are constants $m$ and $n$ such that $\tr E_i^2=m$ for all $i$ and $\tr E_i E_j=n$ for all $i\ne j$.  Since
\be
d=\tr I^2=\sum_{ij}\tr E_i E_j = \sum_i \tr E_i^2 + \sum_{i\ne j} \tr E_i E_j\;,
\ee
$m$ and $n$ must be related by
\be
m+(d^2-1)n=\frac{1}{d}\;.
\ee
On the other hand, with these conditions fulfilled the $E_i$ must all have the same trace.  For,
\be
\tr E_k = \sum_i \tr E_k E_i = m + (d^2-1)n\;.
\ee
Consequently
\be
\tr E_k = \frac{1}{d}\;.
\ee
Now, how large can $m$ be?  Take a positive semi-definite matrix $A$ with $\tr A=1$ and eigenvalues $\lambda_i$.  Then $\lambda_i\le 1$, and clearly $\tr A^2\le \tr A$ with equality if and only if the largest $\lambda_i$ is equal to 1.  Hence, $d E_k$ will give the largest allowed value $m$ if
\be
E_i = \frac{1}{d} \Pi_i = \frac{1}{d} |\psi_i\rangle\langle \psi_i|\;,
\ee
for some rank-1 projection operator $\Pi_i$.  If this obtains, $n$ must follow suit with
\be
n=\frac{1}{d^2(d+1)}\;.
\ee

In total we have shown that a measurement $\{E_i\}$, $i=1,\ldots,d^2$, will achieve the best lower bound of this Schwarz method for $F$ if and only if
\be
E_i=\frac{1}{d} \Pi_i \qquad \mbox{with} \qquad \tr \,\Pi_i\Pi_j=\frac{d\delta_{ij}+1}{d+1}\;.
\label{Mustard}
\ee

Significantly, it turns out that measurements of this variety also have the nice property of being necessarily informationally complete \cite{Caves99}.  Let us show this for completeness:  It is just a matter of proving that the $E_i$ are linearly independent.  Suppose there are some numbers $\alpha_i$ such that
\be
\sum_i \alpha_i E_i = 0\;.
\label{PenScript}
\ee
Taking the trace of this equation, we infer that $\sum_i \alpha_i = 0$. Now multiply Eq.~(\ref{PenScript}) by an arbitrary $E_k$ and take the trace of the result.  We get,
\be
\frac{1}{d^2}\sum_i \alpha_i \frac{d\delta_{ik}+1}{d+1}=0\;.
\ee
In other words
\be
\sum_i \alpha_i \delta_{ik} = 0\;,
\ee
which of course implies $\alpha_k=0$.  So the $E_i$ are linearly independent.

These kinds of measurements are presently a hot topic of study in quantum information theory, and have come to be known as ``symmetric informationally complete'' quantum measurements \cite{Caves99}.  As such, the measurement $\{E_i\}$, the associated set of projection operators $\{\Pi_i\}$, and even the set of $\{|\psi_i\rangle\}$ are often simply called SIC (pronounced ``seek'').\footnote{This choice of pronunciation is meant to be in accord with the pedant's pronunciation of the Latin adverb {\it sic} \cite{Bennett08}.  Moreover, it alleviates any potential confusion between the pluralized form SICs and the number six in conversation.}  We will adopt that terminology here.

Here is an example of a SIC in dimension-3.  Taking $\omega=e^{2\pi i/3}$ to be a third-root of unity and $\overline{\omega}$ to be its complex conjugate, let
\bea
|\psi_1\rangle=\frac{1}{\sqrt2}{\veec110}, &
|\psi_2\rangle=\frac{1}{\sqrt2}{\veec011}, &
|\psi_3\rangle=\frac{1}{\sqrt2}{\veec101},
\nonumber\\
|\psi_4\rangle=\frac{1}{\sqrt2}{\veec1\omega 0}, &
|\psi_5\rangle=\frac{1}{\sqrt2}{\veec01\omega}, &
|\psi_6\rangle=\frac{1}{\sqrt2}{\veec\omega 01},\rule{0mm}{10mm}
\nonumber\\
|\psi_7\rangle=\frac{1}{\sqrt2}{\veec1{\overline{\omega}}0}, &
|\psi_8\rangle=\frac{1}{\sqrt2}{\veec01{\overline{\omega}}}, &
|\psi_9\rangle=\frac{1}{\sqrt2}{\veec{{\overline{\omega}}}01}.\rule{0mm}{10mm}
\label{Moutard}
\eea
One can check by quick inspection that this set of vectors satisfies Eq.~(\ref{Mustard}).

Do SICs exist for every finite dimension $d$?  Despite much effort \cite{Caves99,Zauner99,Renes04,Fuchs04b,Appleby05,ApplebyDangFuchs,Grassl04,%
Grassl05,Grassl06,Grassl08,Grassl08b,Scott06,Appleby06,Klappenecker05,%
Wootters06,Colin05,Flammia06,Kim06,Bos07,Khatirinejad07,Albouy07,Godsil08,%
Bengtsson08,OpenProblemsPage}, no one presently knows.  However, there is a strong feeling in the community that they do, as analytical proofs have been obtained for all dimensions $d=2$--$15$ and 19, and within a numerical precision of $10^{-38}$, they have been observed by computational means \cite{Scott09} in all dimensions $d=2$--$67$.  From here out, we will proceed as if SICs do indeed always exist.  The structure is crucial for revealing the essence of the Born Rule as an addition to coherence.

Let us spell out in some detail what the set of quantum states written as SIC probability vectors $\|p\drangle$ looks like. Perhaps the most remarkable thing about a SIC is the level of simplicity it lends to Eq.~(\ref{YakBoy}).  On top of the theoretical justification that SICs are as near as possible to an orthonormal basis, simply working through the inversion of  $\|p\drangle$ to $\rho$, one gets this beautiful expression \cite{Fuchs04b,Caves02b}:
\be
\rho=\sum_i \left((d+1)p(i)-\frac{1}{d}\right)\Pi_i\;.
\label{FierceUrgencyOfNow}
\ee
In other words, effectively all explicit reference to the matrix $M^{-1}$ disappears from the expression, and one is left with a universal scalar readjustment to the components $p(i)$.  This will have important implications.

Still, one cannot put just any $p(i)$ into Eq.~(\ref{FierceUrgencyOfNow}) and expect to obtain a positive semi-definite $\rho$.  Only some probability vectors $\|p\drangle$ are {\it valid\/} ones.  Which ones are they?  For instance, $p(i)\le \frac{1}{d}$ must be the case, as dictated by Eq.~(\ref{Mustard}), and this already restricts the class of valid probability assignments.  But there are more requirements than that.

In preparation for expressing the set of valid probability vectors $\|p\drangle$, let us note that since the $\Pi_k$ form a basis on the space of operators, we can define the very notion of operator-multiplication in terms of them.  This is done by introducing the so-called ``structure coefficients'' $\alpha_{ijk}$ for the algebra:
\be
\Pi_i\Pi_j=\sum_k \alpha_{ijk}\Pi_k\;.
\label{Hermeneutic}
\ee
A couple of properties follow immediately.  Taking the trace of both sides of Eq.~(\ref{Hermeneutic}), one has
\be
\sum_k \alpha_{ijk} = \frac{d\,\delta_{ij}+1}{d+1}\;.
\label{Fritsa}
\ee
Using this, one gets straightforwardly that
\be
\tr\!\Big( \Pi_i\Pi_j\Pi_k\Big)=\frac{1}{d+1}\left(d\,\alpha_{ijk}+
\frac{d\,\delta_{ij}+1}{d+1}\right).
\ee
In other words,
\be
\alpha_{ijk}=\frac{1}{d}\left((d+1)\tr\!\Big(\Pi_i\Pi_j\Pi_k\Big)-
\frac{d\,\delta_{ij}+1}{d+1}\right).
\ee
For the analogue of Eq.~(\ref{Fritsa}) but with summation over the first or second index, one gets,
\be
\sum_i \alpha_{ijk}=d\delta_{jk} \qquad\mbox{and}\qquad\sum_j \alpha_{ijk}=d\delta_{ik}\;.
\ee

With these expressions in hand, one sees a very direct connection between the structure of the algebra of quantum states when written in operator language and the structure of quantum states when written in probability-vector language.  For, the complete convex set of quantum states is fixed by the set of its extreme points, i.e., the pure quantum states---rank-1 projection operators.  To characterize this set algebraically, one method is to note that these are the only hermitian operators satisfying $\rho^2=\rho$.  Using Eq.~(\ref{FierceUrgencyOfNow}), we find that a quantum state $\|p\drangle$ is pure if and only if its components satisfy these $d^2$ simultaneous quadratic equations:
\be
p(k)=\frac{1}{3}(d+1)\sum_{ij} \alpha_{ijk}\,p(i)p(j)+\frac{2}{3d(d+1)}\;.
\label{NewFavoredBoy}
\ee
Another way to characterize this algebraic variety is to make use of the remarkable theorem of Flammia, Jones, and Linden \cite{Flammia04,Jones05}:  A hermitian operator $A$ is a rank-one projection operator if and only if $\tr A^2=\tr A^3=1$.\footnote{The theorem is nearly trivial to prove once one's attention is called to it:  Since $A$ is hermitian, it has a real eigenvalue spectrum $\lambda_i$. \ From the first condition, one has that $\sum_i \lambda_i^2=1$; from the second, $\sum_i \lambda_i^3=1$. \ The first condition, however, implies that $|\lambda_i| \le 1$ for all $i$. \ Consequently $1-\lambda_i\ge0$ for all $i$.  Now taking the difference of the two conditions, one sees that $\sum_i \lambda_i^2(1-\lambda_i)=0$. \ In order for this to obtain, it must be the case that $\lambda_i$ is always $0$ or $1$ exclusively. That there is only one nonzero eigenvalue then follows from using the first condition again.  Thus the theorem is proved.  Nonetheless, we call this a remarkable theorem with an eye toward its depth as a statement about quantum mechanics, and maybe more importantly because of the sheer number of good quantum theorists who had not known the theorem before it was called to their attention. CAF knows this because of personal interviews with C. M. Caves, D. Gottesman, A. S. Holevo, R. Jozsa, S. Kochen, E. Lieb, G. Lindblad, M. A. Nielsen, A. Uhlmann, and several others.  It is an interesting historical question as to how this little theorem was somehow missed over the years.}$^,$\footnote{The importance of $\tr \rho^2$ as a measure of the information possessed by a quantum state has been especially emphasized in \cite{Brukner01}.  But what is the significance of the cubic entropy $\tr \rho^3$?  This seems to be a new consideration, so far little explored.  Is the quantity only a mathematically convenient tool for characterizing pure states, or is it perhaps the hint of a deeper consideration arising from information theory?  No answers here, but it is an intriguing question.} \ So in fact our $d^2$ simultaneous quadratic equations reduce to just two equations instead, one a  quadratic and one a cubic:
\be
\sum_i p(i)^2=\frac{2}{d(d+1)}
\label{PurePurity}
\ee
and
\be
\sum_{ijk} \alpha_{ijk}\,p(i)p(j)p(k)=\frac{4}{d(d+1)^2}\;.
\label{MonkeyFun}
\ee
Note that Eqs.~(\ref{NewFavoredBoy}) and (\ref{MonkeyFun}) are complex equations, but one could symmetrize them and make them purely real if one wanted to.

There are also some advantages to working out these equations more explicitly in terms of the completely symmetric 3-index tensor
\be
c_{ijk}=\mbox{Re}\;\tr\!\Big(\Pi_i\Pi_j\Pi_k\Big)\;.
\label{TripleSec}
\ee
In terms of these quantities, the analogues of Eqs.~(\ref{NewFavoredBoy}) and (\ref{MonkeyFun}) become
\be
p(k)=\frac{(d+1)^2}{3d}\sum_{ij} c_{ijk}\,p(i)p(j)-\frac{1}{3d}
\label{MegaMorph}
\ee
and
\be
\sum_{ijk} c_{ijk}\,p(i)p(j)p(k)=\frac{d+7}{(d+1)^3}\;,
\label{ChocolateMouse}
\ee
respectively.  The reason for noting this comes from the surprising simplicity of the $d$ matrices $C_k$ with matrix entries $(C_k)_{ij}=c_{ijk}$ from Eq.~(\ref{TripleSec}).  For instance, for each value of $k$, $C_k$ turns out to have the form \cite{Appleby09}
\be
C_k=\|m_k\drangle\dlangle m_k\|+\frac{d}{2(d+1)} Q_k\;,
\label{Magma}
\ee
where the $k$-th vector $\|m_k\drangle$ is defined by
\be
\|m_k\drangle = \left( \frac{1}{d+1}, \ldots, 1, \ldots, \frac{1}{d+1}\right)^{\!\rm T}\;,
\ee
and $Q_k$ is a $(2d-2)$-dimensional projection operator on the real vector space embedding the probability simplex $\Delta_{d^2}$.  Furthermore, using this, one obtains a very pretty expression indeed for the pure states; they are probabilities satisfying a simple class of quadratic equations
\be
p(k)=d\,p(k)^2 + \frac{1}{2}(d+1)\dlangle p\|Q_k\|p\drangle\;.
\ee
Exploring these matrices more thoroughly, however, is beyond the scope of the present paper.  Details will soon be found in Ref.~\cite{Appleby09}.

With Eqs.~(\ref{PurePurity}), (\ref{MegaMorph}), and (\ref {ChocolateMouse}) we have now discussed the extreme points of the convex set of quantum states---the pure states.  The remainder of the set of quantum states is then constructed by taking convex combinations of the pure states. This is an implicit expression of quantum-state space.  But SICs can also help give an explicit parameterization of the convex set.

We can see this by starting not with density operators, but with ``square roots'' of density operators.  This is useful because a matrix $\rho$ is positive semi-definite if and only if it can be written as $\rho=B^2$ for some hermitian $B$.  Thus, let
\be
B=\sum_i b_i\Pi_i
\ee
with $b_i$ a set of real numbers.  Then,
\be
\rho=\sum_k\left(\sum_{ij}b_i b_j\alpha_{ijk}\right)\Pi_k
\ee
will represent a density operator so long as $\tr \rho=1$. This condition requires simply that
\be
\left(\sum_i b_i\right)^2+d\sum_i b_i^2\;=\;d+1\;,
\label{QuadraticVariety}
\ee
so that the vectors $(b_1, \ldots, b_{d^2})$ lie on the surface of an ellipsoid.

Putting these ingredients together with the definition $p(k)=\tr\rho E_k$, we have the following parameterization of valid probability vectors $\|p\drangle$:
\be
p(k)=\frac{1}{d}\sum_{ij}c_{ijk}b_i b_j\;.
\label{Shape}
\ee
Here the $c_{ijk}$ are the triple-product constants defined in Eq.~(\ref{TripleSec}) and the $b_i$ satisfy the constraint (\ref{QuadraticVariety}).
%If one prefers, one could make use of the structure constants %$\alpha_{ijk}$ themselves and alternately write
%\be
%p(k)=\frac{1}{d+1}\sum_{ij}b_i b_j\alpha_{ijk}+\frac{1}{d(d+1)}\;.
%\ee

Finally, let us note what the Hilbert-Schmidt inner product of two quantum states looks like in SIC terms.  If a quantum state $\rho$ is mapped to $\|p\drangle$ via a SIC, and a quantum state $\sigma$ is mapped to $\|q\drangle$, then
\bea
\tr\rho\sigma &=& d(d+1)\sum_i p(i)q(i)-1
\nonumber\\
&=& d(d+1)\dlangle p\|q\drangle - 1\;.
\label{Bibjangles}
\eea
Notice a particular consequence of this:  Since $\tr\rho\sigma\ge0$, the distributions associated with distinct quantum states can never be too nonoverlapping:
\be
\dlangle p\|q\drangle \ge \frac{1}{d(d+1)}\;.
\ee
This will certainly be significant in our later development.

With this development we have given a broad outline of the shape of quantum-state space in SIC terms.  We do this because that shape is our target.  Particularly, we are obliged to answer the following question:  If one takes the view that quantum states are {\it nothing more\/} than probability distributions with the restrictions (\ref{Shape}) and (\ref{QuadraticVariety}), what on earth could motivate that restriction?  That is, what could motivate it {\it other than\/} knowing the usual formalism for quantum mechanics?  The answer has to do with rewriting the Born Rule in terms of SICs, which we will do in Section \ref{KnuckleFinger}.

\subsection{Aside on Unitarity}

Let us take a moment to move beyond statics and rewrite quantum dynamics in SIC terms before moving on:  We do this because the result will have an uncanny resemblance to the Born Rule itself, once developed in the next section.

Suppose we start with a density operator $\rho$ and let it evolve under unitary time evolution to a new density operator $\sigma=U\rho U^\dagger$.  If $\rho$ has a representation $p(i)$ with respect to a certain given SIC, $\sigma$ will have a representation as well---let us call it $q(j)$.  We use the different index $j$ (contrasting with $i$) to help indicate that we are talking about the quantum system at a later time than the original.

What is the form of the mapping that takes $\|p\drangle$ to $\|q\drangle$?  It is simple enough to find with the help of Eq.~(\ref{FierceUrgencyOfNow}):
\be
q(j)=\frac{1}{d}\tr\sigma\Pi_j=\frac{1}{d}\sum_i \left((d+1)p(i)-\frac{1}{d}\right)\tr\!\Big(U\Pi_iU^\dagger\Pi_j\Big)\;.
\ee
Now simply note:  If we define
\be
r(j|i)=\frac{1}{d}\tr\!\Big(U\Pi_iU^\dagger\Pi_j\Big)
\ee
we will have a matrix of quantities such that $0\le r(j|i)\le 1$ and
\be
\sum_j r(j|i) = 1 \quad \forall i \qquad\qquad\mbox{and}\qquad\qquad \sum_i r(j|i) = 1 \quad \forall j\;.
\ee
In other words, the $d^2\times d^2$ matrix $[\,r(j|i)\,]$ is a doubly stochastic matrix \cite{Horn85}.

Most importantly, one has
\be
q(j)=(d+1)\sum_{i=1}^{d^2} p(i) r(j|i) - \frac{1}{d}\;.
\label{NoAccident}
\ee
In other words, unitary time evolution in a SIC representation is amazingly close to simple stochastic evolution!  Or at least formally so.  Surely this teaches us something about unitarity and its connection to the Born Rule itself, as we shall shortly see.

\section{Expressing the Born Rule in Terms of SICs}
\label{KnuckleFinger}

\begin{figure}
\begin{center}
\includegraphics[height=3.8in]{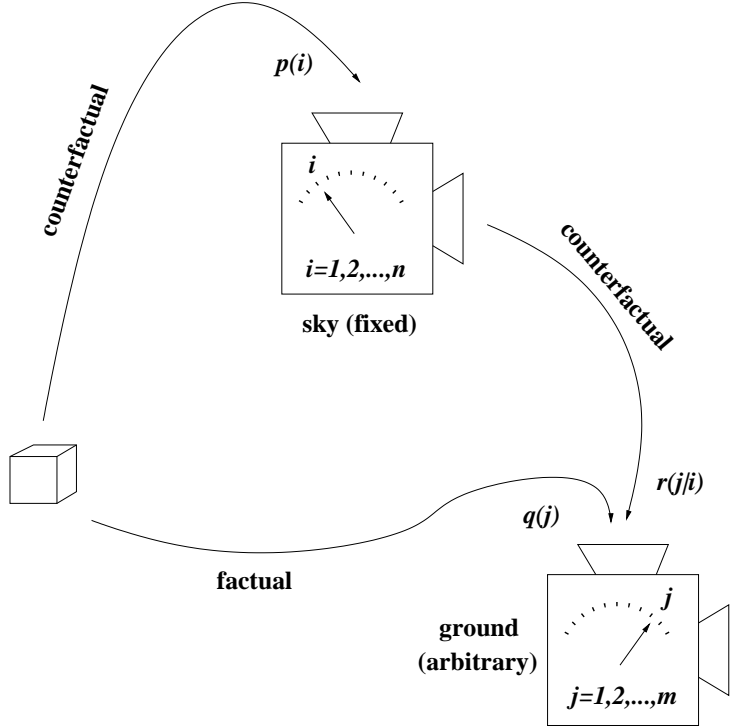}
\bigskip\caption{The diagram above expresses the basic conceptual apparatus of this paper.  The measurement on the ground, with outcomes $j=1,\ldots,m$, is some potential measurement that could be performed in the laboratory---i.e., one that could be factualized.  The measurement in the sky, on the other hand, with outcomes $i=1,\ldots,n$, is a fixed measurement one can contemplate independently.  The probability distributions $p(i)$ and $r(j|i)$ represent how an agent would gamble if a conditional lottery based on the measurement in the sky were operative.  The probability distribution $q(j)$ represents instead how the agent would gamble on outcomes of the ground measurement if the measurement in the sky and the associated conditional lottery were nullified---i.e., they were to never take place at all.  In the quantum case, the measurement in the sky is a SIC with $n=d^2$ outcomes; the measurement on the ground is any POVM.  In pure Bayesian reasoning, there is no necessity that $q(j)$ be related to $p(i)$ and $r(j|i)$ at all.  In quantum mechanics, however, there is a beautiful and mysterious relation:}
\begin{center}
\parbox{4in}{\boxedeqn{
q(j)=(d+1)\sum_{i=1}^{d^2} p(i) r(j|i) - \frac{1}{d}\sum_{i=1}^{d^2} r(j|i)\;.}}
\end{center}
This equation contains the sum content of the Born Rule.
\end{center}
More dramatically, when the measurement on the ground is a von Neumann measurement with outcomes $P_j=|j\rangle\langle j|$---i.e., the usual venue for quantum foundations discussions---then,
$$
q(j)=(d+1)\sum_{i=1}^{d^2} p(i) r(j|i) - 1\;.
$$
\end{figure}

In this section we come to the heart of the paper: We rewrite the Born Rule in terms of SICs.  It is easy enough; we just use the expansion in Eq.~(\ref{FierceUrgencyOfNow}).  Let us first do it for an {\it arbitrary\/} von Neumann measurement---that is, any measurement specified by a set of rank-1 projection operators $P_j=|j\rangle\langle j|$, $j=1,\ldots,d$.  Expressing the Born Rule the usual way, we obtain these probabilities for the measurement outcomes:
\be
q(j)=\tr\rho P_j\,\;.
\ee
Then, by defining
\be
r(j|i)=\tr \Pi_i P_j\;,
\label{Amoebus}
\ee
one sees that the Born Rule becomes
\be
q(j)=(d+1)\sum_{i=1}^{d^2} p(i) r(j|i) - 1\;.
\label{PrettyBoyFloyd}
\ee

Let us take a moment to marvel at the simplicity of this equation and seek out a good interpretation of it.  It feels like one could not ask for a more beautiful expression of the considerations laid out in Section 2.  For imagine that before performing the $P_j$ measurement---we will call it the ``measurement on the ground''---we were to perform a SIC measurement $\Pi_i$.  We will call the latter the ``measurement in the sky.''

Starting with an initial quantum state $\rho$, we would assign a probability distribution $p(i)$ to the outcomes of the SIC measurement.  Moreover, under the assumption that the SIC device leads to a L\"uders Rule conditionalization for the posterior quantum state---so that $\rho$ transforms to $\Pi_i$ when outcome $i$ occurs---the conditional probability for $j$ consequent upon $i$ would be precisely $r(j|i)$ as defined in Eq.~(\ref{Amoebus}).  With these assignments, Dutch-book coherence would then demand an assignment $s(j)$ for the outcomes on the ground that satisfies
\be
s(j)=\sum_{i=1}^{d^2} p(i) r(j|i)\;,
\ee
i.e., a probability that comes about via the Law of Total Probability.

But now imagine the measurement in the sky nullified---that is, imagine it does not occur after all---and that the quantum system goes directly to the measurement device on the ground.  Quantum mechanics tells us to make the probability assignment $q(j)$ given in Eq.~(\ref{PrettyBoyFloyd}) instead.  So,
\be
q(j)=(d+1)s(j)-1
\label{MorningAfter}
\ee
in line with the style of expression indicated in Eq.~(\ref{Moufa!}).  That $\|q\drangle\ne \|s\drangle$ holds, regardless of the assignment of $\|s\drangle$, is a formal expression of the idea that the ``unperformed SIC had no outcomes.'' But Eq.~(\ref{MorningAfter}) tells us still more detailed information than this. It seems to be a kind of ``coherence-plus''---it's not just your old coherence, but rather has a kind of ``value added.''  It tells us something, and we suspect something deep, about the very structure of quantum mechanics.

To support this, let us try to glean some insight from Eq.~(\ref{PrettyBoyFloyd}).  The most obvious thing one can note is that $\|s\drangle$ cannot be too sharp of a probability distribution.  For otherwise $q(j)$ will violate the bounds $0\le q(j)\le 1$ set by Dutch-book coherence.  Particularly,
\be
\frac{1}{d+1}\le s(j) \le \frac{2}{d+1}\;.
\ee
This in turn will have implications for the range of values possible for $p(i)$ and $r(j|i)$.  Indeed if either of these distributions become too sharp (in the latter case, for too many values of $i$), again the bounds will be violated.  Thus, one starts to wonder whether it might just be the case that the essential part of quantum-state space structure, as expressed by its extreme points satisfying Eqs.~(\ref{PurePurity}) and (\ref{MonkeyFun}), arises from the very requirement that $q(j)$ be a proper probability distribution.  In the next Section, we will explore this question in greater depth.

First though, we must note the most general form of the Born Rule. That is to say, general in the sense of the cases when the measurement on the ground is not restricted to being of the simple von Neumann variety.  So, let $q(j)=\tr \rho F_j$ and $r(j|i)=\tr \Pi_i F_j$ for some general positive-operator valued measure $\{F_j\}$ with any number of outcomes, $j=1,\ldots,m$.  Then the Born Rule becomes
\be
q(j)=(d+1)\sum_{i=1}^{d^2} p(i) r(j|i) - \frac{1}{d}\sum_{i=1}^{d^2} r(j|i)\;.
\label{PrettyBoyBabyThisBeLloyd}
\ee
As stated, this is the most general form of the Quantum Law of Total Probability, but even this generalization can have its special cases.  For instance when the measurement on the ground is itself another SIC (any SIC) it reduces to
\be
q(j)=(d+1)\sum_{i=1}^{d^2} p(i) r(j|i) - \frac{1}{d}\;.
\ee
Notice the formal resemblance between this and Eq.~(\ref{NoAccident}) expressing unitary time evolution.

Equation (\ref{PrettyBoyBabyThisBeLloyd}) is not quite as pretty as Eq.~(\ref{PrettyBoyFloyd}), but it is an example of the structure indicated in Eq.~(\ref{Noufa!}).  Indeed, it is of a still more particular form,
\be
q(j)=G\!\left(\sum_i p(i)r(j|i)\, ,\,\sum_i r(j|i)\right).
\ee
In this case, the function $G$ only depends on two terms, a term comprising the classical Law of Total Probability, and a term dependent only upon the sum of the conditional probabilities.

\subsection{Why Complex Hilbert Space, Rather Than Real?}
\label{Vivilify}

Here we take an aside to indicate the necessity of complex Hilbert spaces for our considera\-tions---at least contrasting the complex case with real Hilbert spaces (leaving the question of quaternions for future work).  We start by instigating a discussion of what is actually required to get an equation of {\it the form\/} of Eq.~(\ref{PrettyBoyBabyThisBeLloyd}).  By this, we mean an equation of the form,
\be
q(j)=\alpha\sum_{i=1}^n p(i)r(j|i) - \beta \sum_{i=1}^n r(j|i)\;,
\label{BruceCockburn}
\ee
where $\alpha$, $\beta$, and $n$ need not have the same values as in Eq.~(\ref{PrettyBoyBabyThisBeLloyd}).  In other words, what really is powering the general form of our Quantum Law of Total Probability?

Up till now, we have portrayed this equation as solely a consequence of choosing a SIC for the measurement in the sky.  Indeed this is true, insofar as one limits oneself to a ``minimal'' informationally complete POVM for the sky \cite{Fuchs04b}.  What we mean by a minimal informationally complete POVM is that it have precisely $n=d^2$ outcomes (just enough to span the space of operators).

But what if we back off from the requirement of minimality?  Can we still get our desired equation?  The answer is actually yes:  Any quantum two-design \cite{Zauner99} will do.  A quantum two-design is any set of rank-1 projection operators $P_i$, $i=1,\ldots,n$ such that
\be
\frac{1}{n}\sum_i P_i\otimes P_i= \int d P_\psi \, P_\psi\otimes P_\psi\,,
\ee
where the right-hand side of this equation represents an integration over the unique unitarily invariant measure on rank-1 projection operators $P_\psi$, sometimes called Haar measure.  Thus, the measurement $P_\psi$ with a continuous infinity of outcomes will certainly do as a standard measurement in the sky.  But, for sufficiently large $n$, (namely $n\approx 2^{O(d^3)}$) finite two-designs can always be constructed as well \cite{Wagner91,Bajnok92}.

Let us illustrate this point for a well-known two-design that has nearly a minimal number of points and can actually be constructed in dimensions $d$ that are power of a prime number (say $d=7$ or $d=3^6$ or $d=83^2$, etc.)  In these dimensions, complete sets of {\it mutually unbiased bases\/} (MUBs) are known to always exist \cite{Wootters89,Bandyopadhyay02}, and when they do, they form a quantum two-design \cite{Klappenecker05a}.  What we mean by this is that there are $d(d+1)$ rank-1 projection operators $\Pi^r_s$, with $r=0,\ldots,d$ and $s=1,\ldots,d$, such that
\bea
\tr\big(\Pi^r_s\Pi^r_t\big) &=& \delta_{st}
\\
\tr\big(\Pi^r_s\Pi^q_t\big) &=& \frac{1}{d} \quad\mbox{when}\quad
r\ne q\;.
\label{fifth}
\eea
In other words, the projection operators naturally fall into $d+1$ orthonormal bases of $d$ elements each; the index $r$ denotes which basis, while $s$ denotes which projection operator within a basis.  Constructing a single POVM with $d(d+1)$ outcomes from these projection operators
\be
E^r_s=\frac{1}{d+1}\Pi^r_s\;,
\ee
one obtains an adequate measurement in the sky.  The reason for this is that one can show \cite{Fuchs04b} that the trace-preserving completely positive map $\Phi$ defined by
\bea
\Phi(X) &=&\sum_{r,s} \frac{1}{d+1}\,\Pi^r_s X \Pi^r_s
\nonumber\\
&=&\sum_{r,s} \tr\!\Big(X E^r_s\Big)\/\Pi^r_s\;,
\eea
has a more direct representation as
\be
\Phi(X)=\frac{1}{d+1}\Big((\tr X)I+ X\Big)\;.
\label{HumVee}
\ee
Substituting a quantum state $\rho$ for $X$, the result $\Phi(\rho)$
represents the quantum state an agent would write down to determine his probabilities for the ground, if the imaginary path through the measurement in the sky had actually obtained.

To match our previous notation, let us define $i$ as $i=r d+s$, and $\Pi_i=\Pi^r_s$ and $E_i=\frac{1}{d+1}\Pi_i$. Then $q(j)$, $p(i)$, and $r(j|i)$ come about in analogous fashion to previously (i.e., via the conceptual apparatus of Figure 2), and we get
\be
q(j)=(d+1)\sum_{i=1}^{d(d+1)} p(i) r(j|i) - \frac{1}{d+1}\sum_{i=1}^{d(d+1)} r(j|i)\;.
\label{Nubility}
\ee
The only things that differ between this and the previous Quantum Law of Total Probability, Eq.~(\ref{PrettyBoyBabyThisBeLloyd}), are the range of summation for $i$ and the $\frac{1}{d+1}$ prefixing the second sum instead of $\frac{1}{d}$.

Equation (\ref{Nubility}) is, as promised, of the form  Eq.~(\ref{BruceCockburn}).  What powered this result is a combination of three factors.  1) The $E_i$ are rank-1 operators, so that the states emerging from the device in the sky via L\"uders rule
\be
\rho_i=\frac{1}{\tr\rho E_i}\sqrt{E_i}\rho\sqrt{E_i}
\ee
have no dependence on $\rho$ itself. 2) The weight of each $E_i$, i.e., $w_i=\tr E_i$, is of constant value.  Otherwise the rightmost term in Eq.~(\ref{Nubility}) would have looked like this $\sum_i w_i r(j|i)$, involving a set of extraneous numbers beyond the $r(j|i)$.  And 3)  $\Phi$'s action turned out to be given by Eq.~(\ref{HumVee}), so that for any quantum state $\rho$, $\Phi(\rho)$ contains an unadulterated remnant of $\rho$ itself---namely, $\Phi(\rho)$ is just a mixture of $\rho$ with the identity.

There are, however, two unpalatable things about this construction.
First, complete sets of MUBs are only known to exist when $d$ is the power of a prime number, and they are generally believed {\it not\/} to exist in other dimensions.  For instance, in $d=6$ no more than three mutually unbiased basis have ever been found, even after quite extensive work \cite{Grassl04,Butterley07,Brierley08,Brierley09}, whereas a total of seven are needed for Eq.~(\ref{HumVee}) to be true. The other problem is that, though the probabilities $p(i)$ live in the $n=d(d+1)$-simplex $\Delta_n$, they only inhabit a $d^2$-dimensional subspace of it \cite{Peres98}.  This is a kind of inefficiency in the representation that is surely ugly.  Furthermore, this particular problem is inherited by any quantum two-design with more than a minimal number of elements, becoming acute for the continuous case where the probability simplex is infinite dimensional but only a finite dimensional subspace of it is actually occupied.

One of the very nice things about SICs is that they {\it do seem to\/} exist in all finite dimensions, and thus they form a standard for our construction of a Quantum Law of Total Probability.  Moreover they are adequate to the task with a minimal number of POVM elements, creating no inefficiencies in the representation---the state space spans the full simplex.  With this recognized, let us return to the question of what complex Hilbert space has that real Hilbert space does not?

An immediate question is, in the real Hilbert-space case can we find a measurement adequate for the sky, but with a minimal number of POVM elements?  In this venue, quantum states are elements of ${\mathcal S}(\mathbb{R}^d)$, the space of real, symmetric operators.  This is a space with dimensionality $\frac{1}{2}d(d+1)$; consequently a minimal informationally complete POVM must have precisely $\frac{1}{2}d(d+1)$ elements.

Thus suppose we want a map $\Phi: {\mathcal S}(\mathbb{R}^d)\longrightarrow {\mathcal S}(\mathbb{R}^d)$ of this form
\be
\Phi(X)=\frac{1}{n}\sum_i \Pi_i X\Pi_i\;,
\label{Uncle}
\ee
where the $E_i=\frac{1}{n}\Pi_i$ give a minimal informationally complete POVM.  Then we must have
\be
\frac{1}{n}\sum_{i=1}^{\frac{1}{2}d(d+1)} \Pi_i=I\;,
\ee
and consequently $n=\frac{1}{2}(d+1)$.
Furthermore, suppose for all density operators $\rho$, $\Phi$ has the required characteristic already discussed:
\be
\Phi(\rho)=\frac{1}{\alpha}\Big(\rho + n\beta I\Big)\;.
\label{Dell}
\ee
To be trace preserving,
\be
\alpha = n \beta d + 1\;.
\ee

Just to be explicit, if we have characteristics above, from Eq.~(\ref{Uncle}) we have,
\be
\tr \Phi(\rho) E_j =\sum_i p(i)\, \tr \Pi_i E_j = \sum_i p(i) r(j|i)\;,
\ee
and from Eq.~(\ref{Dell}) we have,
\bea
\tr \Phi(\rho) E_j &=& \frac{1}{\alpha}\Big(q(j) + n\beta\, \tr E_j\Big)
\nonumber\\
&=& \frac{1}{\alpha}\Big(q(j) + \beta \sum_i r(j|i) \Big)
\eea
In other words, we have Eq.~(\ref{BruceCockburn}).  Clearly the process can go backwards as well---from Eq.~(\ref{BruceCockburn}) to Eqs.~(\ref{Uncle}) and (\ref{Dell})---since $\{E_j\}$ is an arbitrary POVM.

Are there any operators $\Pi_i$ that can do this?  Here is a way to see that there generally are not.  Let $\Phi$ act on one of the $\Pi_k$:
\be
\frac{1}{n}\sum_i \Pi_i\Pi_k\Pi_i=\frac{1}{\alpha}\Big(\Pi_k+ n\beta I\Big)\;.
\ee
Using $\Pi_i\Pi_k\Pi_i=\tr(\Pi_i\Pi_k)\,\Pi_i$, it follows that we must have
\be
\Big(\alpha-n(1+\beta)\Big)\Pi_k + \sum_{i\ne k}\Big(\alpha\,\tr(\Pi_i\Pi_k)-n\beta\Big)\Pi_i=0\;.
\ee
But the $\Pi_i$ are linearly independent since we are supposing an informationally complete POVM.  Consequently, it is necessary that
\be
\alpha=n(1+\beta)
\ee
and
\be
\tr(\Pi_i\Pi_k)=\frac{n\beta}{\alpha} \qquad \forall i\ne k\;.
\label{Rufus}
\ee
That is, the $E_i$ must form a real vector space version of the SIC-POVM.

But in the real-vector space case it is known that there is generally no set of projection operators spanning ${\mathcal S}(\mathbb{R}^d)$ that also satisfy Eq.~(\ref{Rufus}).  See Refs.~\cite{Lemmens73,Delsarte75} for details.  For instance, when $d=4$, $\frac{1}{2}d(d+1)=10$, but the maximal number of rank-1 projectors satisfying Eq.~(\ref{Rufus}) is 6.  Another example is when $d=15$; then $\frac{1}{2}d(d+1)= 120$, but the maximal number of equiangular projections is 36.

Thus, we are left with the following statement.  In real vector-space quantum mechanics, there is no minimal informationally complete measurement for the sky that will give us a Quantum Law of Total Probability of the type we have been contemplating.

\subsection{Why ``Coherence-Plus'' Instead of Objective Quantum States?}

Let us come back to the main line of our development:  What we are suggesting is that perhaps Eq.~(\ref{PrettyBoyBabyThisBeLloyd}) should be taken as one of the basic axioms of quantum theory, since it provides a way of thinking of the Born Rule as an addition to Dutch-book coherence.  But, one may well ask, what is our problem with the standard way of expressing the Born Rule in the first place?  How is introducing an addition to Dutch-book coherence conceptually any more palatable than introducing objective quantum states or objective probability distributions?  For, if the program is successful, then the demand that $q(j)$ be a proper probability distribution will place necessary restrictions on $p(i)$ and $r(j|i)$.  This---a skeptic would say---is {\it the\/} very sign that one is dealing with objective (or agent-independent) probabilities in the first place.  Why would a personalist Bayesian accept {\it any\/} a priori restrictions on his probability assignments?  And particularly, restrictions supposedly of empirical origin?

The reply is this.  It is true that through an axiom like Eq.~(\ref{PrettyBoyBabyThisBeLloyd}) one gets a restriction on the ranges of the various probabilities one can contemplate holding.  But that restriction in no way diminishes the functional role of prior beliefs in the makings of an agent's particular assignments $p(i)$ and $r(j|i)$.  That is, this addition to coherence preserves the points expressed by Keynes in Section 2 in a way that objective chance cannot.

Take the usual notion of objective chance, as given operational meaning through David Lewis's {\sl Principal Principle\/} \cite{Lewis86a,Lewis86b}.  If an event $A$ has objective chance $\mbox{ch}(A)=x$, then the subjective, personalist probability an agent (any agent) should ascribe to $A$ on the condition of knowing the chance proposition is
\be
\mbox{Prob}\Big(A\;\Big|\;\mbox{``ch}(A)=x\mbox{''}\wedge E\Big)=x
\ee
where $E$ is any ``compatible'' proposition.  There is some debate about what precisely constitutes a compatible proposition, but an example of a proposition universally accepted to be compatible in spite of these interpretive details is this:
$$
E = \mbox{``All my experience causes me to believe $A$ with probability 75\%.''}
$$
That is, upon knowing a chance, all prior beliefs should be overridden. Regardless of the agent's firmly held belief about $A$, that belief should be revised if he is apprised of the objective chance.

When it comes to quantum mechanics, philosophers of science who find something digestible in Lewis's idea, often view the Born Rule itself as a healthy serving of Principal Principle.  Only, it has the quantum state $\rho$ filling the role of chance.  That is, for any agent contemplating performing a measurement $\{P_j\}$, his subjective, personal probabilities for the outcomes $j$ should condition on knowledge of the quantum state just as one conditions with the Principal Principle:
\be
\mbox{Prob}\big(j\;\big|\;\rho \wedge E\big)=\tr\rho P_j\;,
\ee
where $E$ is any ``compatible'' proposition.  Beliefs are beliefs, but quantum states are something else:  They are the facts of nature that power a quantum version of the Principal Principle. In other words, in this context one has conceptually
$$
\rho \quad \longrightarrow \quad \mbox{``ch}(j)=\tr\rho P_j\mbox{''}\;.
$$
Quantum states on this view are the very antithesis of the Keynesian remark---they are not ``part of our human outfit, perhaps given us merely by natural selection.''

But the quantum-Bayesian view cannot abide by this.  For, the essential point for a quantum-Bayesian is that there is no such thing as {\it the\/} quantum state.  There are potentially as many states for a given quantum system as there are agents.  And that point is not diminished by accepting the addition to coherence described in this paper.  Indeed, it is just as with standard (nonquantum) probabilities, where their subjectivity is not diminished by normatively satisfying standard Dutch-book coherence.

The most telling reason for this arises directly from quantum statistical practice.  The way one comes to a quantum-state assignment is ineliminably dependent on one's priors \cite{Fuchs02,Caves07,Fuchs09}.  Quantum states are not god-given, but have to be fought for via measurement, updating, calibration, computation, and any number of related pieces of work.  The only place quantum states are ``given'' outright---that is to say, the model on which much of the notion of an objective quantum state arises from in the first place---is in a textbook homework problem.  For instance, a textbook exercise might read, ``Assume a hydrogen atom in its ground state. Calculate \ldots.''  But outside the textbook it is not difficult to come up with examples where two agents looking at the same data, differing only in their prior beliefs, will asymptotically update to distinct (even orthogonal) pure quantum-state assignments for the same system \cite{Fuchs09}.  Thus the basis for one's particular quantum-state assignment is always {\it outside\/} the formal apparatus of quantum mechanics.\footnote{Nor, does it help to repeat over and over, as one commonly hears coming from the philosophy of physics community, ``quantum probabilities are specified by physical law.''  The simple reply is, ``No, they're not.''  The phrase has no meaning once has taken it on board that quantum states are born in probabilistic considerations, rather than the parent of them, as laboratory practice clearly shows \cite{Paris04,Kaznady09}.}

This is the key difference between the set of ideas being developed here and the dreams of the objectivists: added relations for probabilities, yes, but no one of those probabilities can be objective in the sense being any less a pure function of the agent.  A way to put it more prosaically is that these normative considerations may narrow the agent from the full probability simplex to the set of quantum states, but beyond that, the formal apparatus of quantum theory gives him no guidance on which quantum state he should choose.  Instead, the role of a normative reading of the Born Rule is as it is with usual Dutch book. Here is the way L. J. Savage put it rather eloquently \cite[p.~57]{Savage54}.
\bq
According to the personalistic view, the role of the mathematical theory of probability is to enable the person using it to detect inconsistencies in his own real or envisaged behavior.  It is also understood that, having detected an inconsistency, he will remove it.  An inconsistency is typically removable in many different ways, among which the theory gives no guidance for choosing.
\eq
If an agent does not satisfy Eq.~(\ref{PrettyBoyBabyThisBeLloyd}) with his personal probability assignments, then he is not recognizing properly the change of conditions (or perhaps we could say `context'\footnote{We add this alternative formulation so as to place the discussion within the context of various other analyses of the idea of `contextuality' \cite{Mermin93,Appleby05c,Spekkens05,Spekkens08,Ferrie08}.}) that a potential SIC measurement would bring about.  The theory gives no guidance for which of his probabilities should be adjusted or how, but it does say that they must be adjusted or ``undesirable consequences'' will become unavoidable.

Expanding on this point, Bernardo and Smith put it this way in Ref.~\cite[p.~4]{Bernardo94}:
\bq
Bayesian Statistics offers a rationalist theory of personalistic beliefs in contexts of uncertainty, with the central aim of characterising how an individual should act in order to avoid certain kinds of undesirable behavioural inconsistencies.  \ldots\   The goal, in effect, is to
establish rules and procedures for individuals concerned with
disciplined uncertainty accounting.  The theory is not descriptive, in the sense of claiming to model actual behaviour.  Rather, it is
prescriptive, in the sense of saying `if you wish to avoid the
possibility of these undesirable consequences you must act in the
following way.'
\eq
So much, indeed, we imagine for the {\it full\/} formal structure of quantum mechanics (including dynamics, tensor-product structure, etc.)---that it is all or nearly all an addition to Dutch-book coherence.  And specifying those ``undesirable consequences'' in terms independent of the present considerations is a significant part of the project of specifying the ontology underlying the quantum-Bayesian position. But that is a goal we are not yet prepared to tackle head on.  Instead, let us first explore the consequences of adopting Eq.~(\ref{PrettyBoyBabyThisBeLloyd}) as a basic statement, acting as if we do not yet know the underlying Hilbert-space structure that gave rise to it.

\section{Deriving Quantum-State Space from ``Coherence-Plus''}

\label{Heimlich}

Let us see how far we can go toward deriving various general features of quantum-state space from the conceptual apparatus portrayed in Figure 2.  Particularly, we want to explore how much of the structure represented by the convex hull of either Eq.~(\ref{MegaMorph}) or Eqs.~(\ref{PurePurity}) and (\ref{MonkeyFun}), both thought of as algebraic varieties within the probability simplex $\Delta_{d^2}$ \cite{Sullivant}, can be recovered from these considerations. We will also have to add at least three other assumptions on the nature of quantum measurement, but at first, let us try to forget as much about quantum mechanics as we can.

Namely, start with Figure 2 but forget about quantum mechanics and forget about SICs.  Simply visualize an imaginary experiment in the sky $S$, supplemented with various real experiments we might perform on the ground $G$.  We postulate that the probabilities we should ascribe for the outcomes of $G$, are determined by the probabilities we would ascribe to the imaginary outcomes in the sky and the conditional probabilities for the outcomes of $G$ consequent upon them, were the measurement in the sky real.  Particularly we postulate
\be
q(j)=(d+1)\sum_{i=1}^{d^2} p(i) r(j|i) - \frac{1}{d}\sum_{i=1}^{d^2} r(j|i)\;.
\label{WaitingForBread}
\ee
We call this postulate the {\it \textbf{urgleichung}} to emphasize its primality.  As before, $p(i)$ represents the probabilities in the sky and $q(j)$ represents the probabilities on the ground.  The index $i$ is assumed to range from 1 to $d^2$, for some fixed natural number $d$.  The range of $j$ will not be fixed, but in any considered case will be denoted as running from 1 to $m$.  (For example, for some cases $m$ might be $d^2$, for some cases it might be $d$, but it need be neither and may be something else entirely---it will depend upon which experiment we are talking about for $G$.)  We write $r(j|i)$ to represent the conditional probability for obtaining $j$ on the ground, given that $i$ was found in the sky.  When we want to suppress components, we will write vectors $\|p\drangle$ and $\|q\drangle$, and write $R$ for the matrix with entries $r(j|i)$---by definition, $R$ is a stochastic matrix (though not necessarily a doubly stochastic matrix) \cite[pp.~526--528]{Horn85}.

One of the main features we will require, of course, is that calculated by Eq.~(\ref{WaitingForBread}), $\|q\drangle$ must satisfy $0\le q(j)\le 1$ for all $j$.  Thus, let us also honor the special inequality
\be
0\;\le\; (d+1)\sum_{i=1}^{d^2} p(i) r(j|i) - \frac{1}{d}\sum_{i=1}^{d^2} r(j|i)\;\le\; 1
\label{TheTrueUr}
\ee
with a name:  the {\it \textbf{urungleichung}}.

To proceed, let us define two sets $\mathcal P$ and $\mathcal R$, the first consisting of priors for the sky $\|p\drangle$, and the second consisting of stochastic matrices $R$.  We shall sometimes call $\mathcal P$ our {\it state space}, and its elements {\it states}.
We will say that $\mathcal P$ and $\mathcal R$ are {\it consistent\/} (with respect to the urungleichung) if 1) for any fixed $R\in \mathcal R$, there is no $\|p\drangle\in\mathcal P$ that does not satisfy the urungleichung, and 2) for any fixed $\|p\drangle\in\mathcal P$, there is no $R\in \mathcal R$ that does not satisfy the urungleichung.  With respect to consistent sets $\mathcal P$ and $\mathcal R$, for convenient terminology, we call a general $\|p\drangle\in\Delta_{d^2}$ {\it valid\/} if it is also within the state space $\mathcal P$; if it is not within $\mathcal P$, we call it invalid.

What we want to pin down are the properties of $\mathcal P$ and $\mathcal R$ under the assumption that they are {\it maximal}.  By this we mean that for any $\|p^\prime\drangle\notin\mathcal P$, if we were to attempt to create a new state space ${\mathcal P}^\prime$ by adding $\|p^\prime\drangle$ to the original $\mathcal P$, ${\mathcal P}^\prime$ and $\mathcal R$ would not be consistent.  Similarly, if we were to attempt to add a new point $R^\prime$ to $\mathcal R$.  In other words, when $\mathcal P$ and $\mathcal R$ are maximal, they are full up with respect to any further additions.  In summary,
\begin{enumerate}
\item
$\mathcal P$ and $\mathcal R$ are said to be {\it consistent\/} if all pairs $\big(\|p\drangle,R\big) \in {\mathcal P}\times {\mathcal R}$ obey the urungleichung.
\item
$\mathcal P$ and $\mathcal R$ are said to be {\it maximal\/} whenever ${\mathcal P}^\prime\supseteq {\mathcal P}$ and ${\mathcal R}^\prime \supseteq {\mathcal R}$ imply ${\mathcal P}^\prime={\mathcal P}$ and ${\mathcal R}^\prime={\mathcal R}$ for any consistent ${\mathcal P}^\prime$ and ${\mathcal R}^\prime$.
\end{enumerate}
There is, of course, no guarantee without further assumptions there will be a unique maximal $\mathcal P$ and $\mathcal R$ consistent with the urungleichung, or even whether there will be a unique set of them up to isomorphism, but we can certainly say some things.

One important result follows immediately:  If $\mathcal P$ and $\mathcal R$ are consistent and maximal, both sets must be convex.  For instance, if $\|p\drangle$ and $\|p^\prime\drangle$ satisfy (\ref{TheTrueUr}) for all $R\in \mathcal R$ it is clear that, for any $x\in [0,1]$, $\|p^{\prime\prime}\drangle=x \|p\drangle + (1-x) \|p^\prime\drangle$ will as well.  Thus, if $\|p^{\prime\prime}\drangle$ were not in $\mathcal P$, it would not have been maximal to begin with.\footnote{It is important to recognize that the considerations leading to the convexity of the state space here are a distinct species from the arguments one finds in the ``convex sets'' and ``operational'' approaches to quantum theory. See for instance \cite{Holevo82,Busch95} and more recently the BBLW school starting in Ref.~\cite{Barnum06} (and several publications thereafter), as well as the work of Hardy \cite{Hardy01}.  There the emphasis is on the idea that a state of ignorance about a finer preparation is a preparation itself.  The present argument is even different than some of our own earlier Bayesian considerations (where care was taken {\it not\/} to view `preparation' as an objective matter of fact, independent of prior beliefs, as talk of preparation would seem to imply) \cite{Fuchs02,Schack04}.  Here instead, the emphasis is on the closure of the urungleichung.  The philosophy behind this is the strict priority of Eq.~(\ref{WaitingForBread}) as a ``law of thought'' in the quantum context: So long as a usage of the urgleichung generates no troubles in its application, why not include that usage as a valid one? The focus on maximal $\mathcal P$ and $\mathcal R$ mathematizes this basic idea.}

Now, is there any obvious connection between $\mathcal P$ and $\mathcal R$?  Let us make the completely innocuous assumption that one can be completely ignorant of the outcomes in the sky, so that
\be
\|p\drangle=\left(\frac{1}{d^2},\frac{1}{d^2},\ldots,\frac{1}{d^2}\right)^{\!\rm T}\in {\mathcal P}\;.
\ee
Certainly for any real-world experiment, one can always be maximally ignorant of which of its outcomes will occur! Suppose now that the experiment in the sky really is performed as well as the experiment on the ground, but that the agent only comes to know the outcome $j$ on the ground, remaining ignorant of the outcome in the sky.  Then the outcome on the ground will teach the agent something about the outcome in the sky.  Using Bayes rule to invert probabilities, he will have
\be
\mbox{Prob}(i|j)=\frac{r(j|i)}{\sum_k r(j|k)}\;.
\label{NoseHair}
\ee
Let us now make a less innocuous assumption:
\begin{assump}{\rm $\!\!$:}
{\rm Principle of Reciprocity:\ Posteriors Are Priors.}  For any $R\in\mathcal R$, a posterior probability $\mbox{\rm Prob}(i|j)$ as in Eq.~(\ref{NoseHair}) is a valid prior $p(i)$ for the outcomes of the measurement in the sky. Moreover, all valid priors $p(i)$ may arise in this way (though one may have other reasons for adopting the prior than this).
\end{assump}
Quantum mechanics certainly has this property.  For, suppose a completely mixed state for our quantum system and a POVM ${\mathcal G}=\{G_j\}$ measured on the ground.  Upon noting an outcome $j$ on the ground, the agent will use Bayes' Rule to infer
\be
\mbox{Prob}(i|j)=\frac{\tr\Pi_i G_j}{d\,\tr G_j}\;.
\ee
Defining
\be
\rho_j=\frac{G_j}{\tr G_j}\;,
\ee
this says that
\be
\mbox{Prob}(i|j)=\frac{1}{d}\,\tr\rho_j\Pi_i\;.
\ee
In other words, $\mbox{Prob}(i|j)$ is itself a SIC-representation of a quantum state.  Moreover, $\rho_j$ can be any quantum state whatsoever, simply by adjusting which POVM $\mathcal G$ is under consideration.  In Copenhagen-interpretation style terminology, this is simply a statement that any quantum state can be ``prepared'' by a suitably chosen measurement device.

It seems to us however that this is a nontrivial assumption generally---particularly when one has fully recognized a situation in which ``unperformed measurements have no outcomes.''  Let us take it nonetheless, saving a better analysis for future work.  What it tells us is that we may think of the set $\mathcal R$ as the primary one, with $\mathcal P$ determined in terms of it.

\subsection{Basis Distributions}

\label{BengalTiger}

Since we are free to contemplate any measurement on the ground, let us consider the case where the ground measurement is set to be the same as that of the sky.  We will denote $r(j|i)$ by $r_{\rm \scriptscriptstyle S}(j|i)$ in this special case.  The urgleichung then requires for any valid $\|p\drangle$ that
\be
p(j)=(d+1)\sum_i p(i)r_{\rm \scriptscriptstyle S}(j|i) - \frac{1}{d} \sum_i r_{\rm \scriptscriptstyle S}(j|i)\;.
\ee
Take the case where $p(i)=\frac{1}{d^2}$ specifically.  This case implies that the $r_{\rm \scriptscriptstyle S}(j|i)$ must satisfy
\be
\sum_i  r_{\rm \scriptscriptstyle S}(j|i)=1\;.
\ee
Therefore, when going back to more general priors $\|p\drangle$, one has in fact the simpler relation
\be
p(j)=(d+1)\sum_i p(i)r_{\rm \scriptscriptstyle S}(j|i) - \frac{1}{d}\;.
\label{FemtoFoot}
\ee
Introducing an appropriately sized matrix $M$ of the form
\be
M=\left(
  \begin{array}{cccc}
    (d+1)-\frac{1}{d} & -\frac{1}{d} & -\frac{1}{d} & -\frac{1}{d} \\
    -\frac{1}{d} & (d+1)-\frac{1}{d} & -\frac{1}{d} & -\frac{1}{d} \\
    -\frac{1}{d} & -\frac{1}{d} & (d+1)-\frac{1}{d} & -\frac{1}{d} \\
    -\frac{1}{d} & -\frac{1}{d} & -\frac{1}{d} & (d+1)-\frac{1}{d} \\
  \end{array}
\right)
\ee
we can rewrite Eq.~(\ref{FemtoFoot}) in vector form,
\be
M R_{\rm \scriptscriptstyle S} \|p\drangle=\|p\drangle\;.
\label{HerbertFoot}
\ee

At this point, we pause for a minor assumption on our state space:
\begin{assump}
The elements $\|p\drangle\in{\mathcal P}$ span the full simplex $\Delta_{d^2}$.
\label{BellyRoar}
\end{assump}
This is a very natural assumption:  For if ${\mathcal P}$ did not span the simplex, one would be justified in simply using a smaller simplex for all considerations.\footnote{Please note how this meshes with the discussion in Section \ref{Vivilify}.  If our starting point were a variation of the Law of Total Probability, but with distributions $\|p\drangle\in P$ with more outcomes than $d^2$, we would have to make more complicated ``subspace assumptions'' of an appropriate variety at this point.}

With Assumption \ref{BellyRoar}, the only way Eq.~(\ref{HerbertFoot}) can be satisfied is if
\be
M R_{\rm \scriptscriptstyle S} = I\;.
\ee
Since $M$ is a circulant matrix, its inverse is a circulant matrix as well, and one can easily work out that,
\be
r_{\rm \scriptscriptstyle S}(j|i)=\frac{1}{d+1}\left(\delta_{ij}+\frac{1}{d}\right)\;.
\ee
It follows by the Principle of Reciprocity (our Assumption 1) then that among the distributions in $\mathcal P$, along with the uniform distribution, there are at least $d^2$ other ones, namely:
\be
\|e_k\drangle= \left(\, \frac{1}{d(d+1)}, \ldots, \frac{1}{d}, \ldots, \frac{1}{d(d+1)}\,\right)^{\!\rm T}\;,
\ee
with a $\frac{1}{d}$ in the $k^{\rm th}$ slot and $\frac{1}{d(d+1)}$ in all other slots.  We shall call these $d^2$ special distributions, appropriately enough, the {\it basis distributions}.

The savvy reader will note that the basis distributions are just the SIC states themselves when expressed in SIC language, only now justified by alternative means.  Particularly, note that
\be
\sum_i e_k(i)^2=\frac{2}{d(d+1)}\quad\forall k\,,
\label{EvaCassidyMoment}
\ee
in accordance with Eq.~(\ref{PurePurity}).

\subsection{A Bloch Sphere}

Consider a class of measurements for the ground that have a property we shall call {\it in-step unpredictability}, ISU.  The property is this:  Whenever one assigns a uniform distribution for the measurement in the sky, one also assigns a uniform distribution for the measurement on the ground. This is meant to express the idea that the measurement on the ground has no in-built bias with respect to one's expectations of the sky: Complete ignorance of the outcomes of one translates into complete ignorance of the outcomes of the other.  (In the full-blown quantum mechanical setting, this corresponds to a POVM with outcomes $G_j$ such that $\tr G_j$ is a constant value---von Neumann measurements with $d$ outcomes being one special case of this.)

Denote the $r(j|i)$ in this special case by $r_{\rm \scriptscriptstyle ISU}(j|i)$, and suppose the measurement being spoken of has $m$ outcomes.  Our demand is that
\be
\frac{1}{m}=\frac{(d+1)}{d^2}\sum_i r_{\rm \scriptscriptstyle ISU}(j|i) - \frac{1}{d} \sum_i r_{\rm \scriptscriptstyle ISU}(j|i)\;.
\ee
To meet this, we must have
\be
\sum_i r_{\rm \scriptscriptstyle ISU}(j|i)=\frac{d^2}{m}\;,
\ee
and the urgleichung becomes
\be
q(j)=(d+1)\sum_i p(i) r_{\rm \scriptscriptstyle ISU}(j|i) - \frac{d}{m}\;.
\label{Lichen}
\ee

Suppose now that a prior $\|s\drangle$ for the sky happens to arise in accordance with Eq.~(\ref{NoseHair}) for one of these ISU measurements.  That is,
\be
s(i)=\frac{r_{\rm\scriptscriptstyle ISU}(j|i)}{\sum_k r_{\rm\scriptscriptstyle ISU}(j|k)}\;,
\label{FlotsamJetsam}
\ee
for some $R_{\rm\scriptscriptstyle ISU}$ and some $j$.
Then Eq.~(\ref{Lichen}) tells us that for any $\|p\drangle\in\mathcal P$, we must have
\be
0\,\le\, \frac{d^2}{m}(d+1)\sum_i p(i) s(i) - \frac{d}{m}\,\le\, 1\;.
\ee
In other words, for any $\|s\drangle$ of our specified variety and any $\|p\drangle\in\mathcal P$, the following constraint must be satisfied
\be
\frac{1}{d(d+1)}\,\le\, \sum_i p(i) s(i) \,\le\, \frac{d+m}{d^2(d+1)}\;.
\ee

Think particularly on the case where $\|s\drangle=\|p\drangle$.  Then we must have
\be
\sum_i p(i)^2 \,\le\, \frac{d+m}{d^2(d+1)}\;.
\label{MommaToldMeNotToCome}
\ee
Note how this compares to Eq.~(\ref{EvaCassidyMoment}).

Further suppose that there are $d^2$ $m$-outcomed measurements with in-step unpredictability, each giving rise (via the posteriors of one of their outcomes) to one of the basis distributions $\|e_k\drangle$.  That is, suppose each basis distribution can be generated as the posterior of an appropriately chosen ISU measurement. (This is a more particular assumption than the Principle of Reciprocity alone can underwrite, so it is certainly adding something new.)  If this is so, then note that according to Eq.~(\ref{EvaCassidyMoment}), the bound in Eq.~(\ref{MommaToldMeNotToCome}) will be violated unless $m\ge d$.  Moreover, it will not be tight for the basis states unless $m=d$ precisely.

Thinking of a basis distribution as the prototype of an extreme-point state (for after all, they give the most predictability possible for the measurement in the sky), this motivates the next assumption---this one being significantly stronger than the previous two:
\begin{assump}
Every extreme point $\|p\drangle\in\mathcal P$ arises as the posterior of an ISU measurement (i.e., in the manner of Eq.~(\ref{FlotsamJetsam})) with $m=d$ and, in fact, achieves equality in Eq.~(\ref{MommaToldMeNotToCome}).
\end{assump}
Thus, for {\it any\/} two extreme points $\|p\drangle$ and $\|s\drangle$, we are assuming
\be
\frac{1}{d(d+1)}\,\le\, \sum_i p(i) s(i) \,\le\, \frac{2}{d(d+1)}\;,
\label{SayHiToYourKnee}
\ee
with equality in the right-hand side when $\|s\drangle=\|p\drangle$.\footnote{By the way, it should be noted that this inequality establishes that if $\mathcal P$ at least contains the actual quantum-state space, it can contain no more than that.  That is, the full set of quantum states is, in fact, a maximal set. For suppose a SIC exists, yet $\|s\drangle$ corresponds to some non-positive-semidefinite operator via the mapping in Eq.~(\ref{FierceUrgencyOfNow}).  Then there will be some $\|p\drangle\in \mathcal P$ corresponding to a pure quantum state such that the left-hand side of Eq.~(\ref{SayHiToYourKnee}) is violated.  This follows immediately from the definition of positive semi-definiteness and the expression for Hilbert-Schmidt inner products in Eq.~(\ref{Bibjangles}).}

Thus, the extreme points of $\mathcal P$ live on a sphere
\be
\sum_i p(i)^2 \,=\, \frac{2}{d(d+1)}\;,
\label{JiggleMeBoog}
\ee
but the requirement of Eq.~(\ref{SayHiToYourKnee}) for any two extreme points implies significantly more structure than that.  The only question is how much?  Certainly, further trivial aspects of quantum-state space follow immediately.  For instance, since the basis distributions are among the set of valid states, for any other valid state $\|p\drangle$ no component in it can be too large.  This follows because
\be
\dlangle p\| e_k\drangle=\frac{1}{d(d+1)}+\frac{1}{d+1}p(k)\;.
\ee
The right-hand side of Eq.~(\ref{SayHiToYourKnee}) then requires
\be
p(k)\le \frac{1}{d}\;.
\ee
But, do we have enough to get to us all the way to Eq.~(\ref{ChocolateMouse}) in addition to Eq.~(\ref{PurePurity})?
We will analyze aspects of this in the next subsection.  First however, let us linger a bit over the significance of the sphere.

What we have postulated in a natural way is that the extreme points of $\mathcal P$ must live on a $(d^2-1)$-sphere centered at the zero vector.  But then it comes for free that these extreme points must also live on a smaller-radius $(d^2-2)$-sphere centered at
\be
\|c\drangle=\left(\frac{1}{d^2}, \frac{1}{d^2}, \ldots, \frac{1}{d^2}\right)^{\!\rm T}\;.
\ee
This is because the $\|p\drangle$ live on the probability simplex $\Delta_{d^2}$.  For, let $\|w\drangle=\|p\drangle-\|c\drangle$, where $\|p\drangle$ is any point satisfying Eq.~(\ref{JiggleMeBoog}).  Then
\be
r^2=\dlangle w\| w\drangle=\frac{d-1}{d^2(d+1)}
\label{SherylCrow}
\ee
gives the radius of the smaller sphere.

The sphere in Eq.~(\ref{SherylCrow}) is actually the more natural sphere for us to think about, as most of the sphere in Eq.~(\ref{JiggleMeBoog})---all but a set of measure zero---is thrown away anyway.  In fact, it may legitimately be considered the higher-dimensional analog of the Bloch sphere from the quantum-Bayesian point of view.  Indeed, when $d=2$, we have a $2$-sphere, and it is isomorphic to the usual Bloch sphere.

It is natural to think of the statement
\be
\sum_i p(i)^2 \,\le\, \frac{2}{d(d+1)} \qquad \mbox{for all $\|p\drangle\in \mathcal P$}
\label{WoohMaBoog!}
\ee
in information theoretic terms.  This is because two well-known measures of the uncertainty associated with a probability assignment---the Renyi and Dar\'oczy entropies \cite{Aczel75} of order 2---are simple functions of the left-hand side of it.  Recall the Renyi entropies most generally (defined for all $\alpha\ge1$)
\be
R_\alpha(\|p\drangle)=\frac{1}{1-\alpha}\ln\!\left(\sum_i p(i)^\alpha\right)
\ee
as well as the Dar\'oczy entropies
\be
D_\alpha(\|p\drangle)=\frac{1}{2^{1-\alpha}-1}\left(\sum_i p(i)^\alpha-1\right)\;.
\ee
In the limit $\alpha\rightarrow 1$, these both converge to the Shannon entropy---which has been their main theoretical interest\footnote{To a casual look, these two entropies might appear to be trivial rescalings of each other. Therefore, though two kinds of entropy have been presented, it would be stylistically better to talk only of one.  This assessment however is incorrect.  The difference between the two particularly comes out when one considers joint probability distributions over more than one variable.  For instance, the Renyi entropies are additive for independent random variables, but not generally subadditive, whereas the Dar\'oczy entropies are always subadditive, just not additive.  See Refs.~\cite{Aczel75,Fuchs96}.}---but from time to time a use arises for some specific values $\alpha\ne1$.  Apparently part of the characterization of quantum-state space is now in this list.

To put it in a slogan \cite{Fuchs01,Caves96,Caves02c}, ``In quantum mechanics, maximal information is not complete and cannot be completed.''  The sharpest predictability one can have for the outcomes of a SIC measurement is specified by Eq.~(\ref{PurePurity}).  This is an old idea, of course, but quantified here in yet another way.  In fact, until the considerations of this paper, it had long been a dream of one of the authors that the majority of the structure of quantum mechanics might come about from a suitably quantified maximal-information-is-not-complete principle.  Take this discussion in Ref.~\cite{Caves96} as an example:
\bq
\indent Formally, one says that in classical physics, maximal information is complete, but in quantum physics, it is not.  What should we demand of a physical theory in which maximal information is not complete?  Maximal information is a state of knowledge; the Bayesian view says that one must assign probabilities based on the maximal information. Classical physics is an example of the special case in which all the resulting probabilities predict unique measurement results; i.e., maximal information is complete.  In a theory where maximal information is not complete, the probabilities one assigns on the basis of maximal information are probabilities for answers to questions one might address to the system, but whose outcomes are not necessarily predictable (some outcomes must be unpredictable, else the maximal information becomes complete).  This implies that the possible outcomes cannot correspond to actualities, existing objectively prior to
asking the question; otherwise, how could one be said to have maximal information?  Furthermore, the theory must provide a rule for assigning probabilities to all such questions; otherwise, how could the theory itself be complete?  Quantum physics is consistent with these demands.
A more ambitious program would investigate whether the quantum rule is the unique rule for assigning probabilities in situations where maximal information is not complete. You won't be surprised to learn that we don't know how to make progress on this ambitious program.
\eq

Part of that ``ambitious program,'' indeed, has recently had some notable success, particularly with the toy model of R.~W. Spekkens \cite{Spekkens07}. In that model, based on local hidden variables and an ``epistemic constraint'' on an agent's knowledge of the variables' values, more than twenty well-known quantum information theoretic phenomena (like no-cloning \cite{Wootters82,Dieks82}, no-broadcasting \cite{Barnum96}, teleportation \cite{Bennett93}, correlation monogamy \cite{Coffman00}, ``nonlocality without entanglement'' \cite{Bennett99}, etc.)\ are readily reproduced, at least in a qualitative way.  But the toy model at the same time reveals a deep flaw in our earlier formulation.  For in it, maximal information is not complete, and yet it is explicitly a hidden-variable model.  This indicates strongly that an information constraint alone cannot support the more sweeping part of the program, that ``the possible outcomes cannot correspond to actualities, existing objectively prior to asking the question.''

The trouble with the phrase ``maximal information is not complete'' and the imagery it entails is that, try as one might to portray it otherwise (by adding ``cannot be completed,'' say), it hints of hidden variables.  What else could the ``not complete'' refer to?  In ways, there is a world of difference between the present considerations to do with an addition to coherence and the raw ``epistemic restriction'' approach advocated previously.  First, it is hard to see how that line of thought can get beyond the possibility of an underlying hidden-variable model (as the toy model amply illustrates).  But second, and more importantly, in the present approach the epistemic restriction comes from a deeper consideration to do with the coherence between factual and counterfactual gambles.  Herein, there is some leeway to fulfill the complementary side of the old dream:  That whatever else the dream incorporates, it should incorporate the idea that unperformed measurements have no outcomes.\footnote{A very old dream, indeed.  Cf.\ the 19 July 1996 letter to S. L. Braunstein  in \cite{Fuchs01}:  ``I don't think there's anything interesting to be gained from {\it simply\/} trying to redo the Coxian `plausibility' argument \cite{Cox46} but with complex numbers. It seems to me that it'll more necessarily be something along the lines of: When you ask me, `Where do all the quantum mechanical outcomes come from?'  I must reply, `There is no where there.'  [With apologies to Gertrude Stein!] That is to say, my favorite `happy' thought is that when we know how to properly take into account the piece of prior information that `there is no where there' concerning the origin of quantum
mechanical measurement outcomes, then we will be left with
`plausibility spaces' that are so restricted as to be isomorphic to
Hilbert spaces.  But that's just thinking my fantasies out loud.''}

The main point is this:  The Bloch sphere may well express an epistemic constraint---a constraint on an agent's advised certainty.  But the epistemic constraint is itself a result of a deeper consideration, not a starting point.  And the constraint is not really expressible in terms of a single information function anyway; instead it involves pairs of distributions.  We go on to explain this point.

\subsection{But Only Part of It}

\label{Blubbery}

The state-space implied by Eq.~(\ref{SayHiToYourKnee}) does not lead to the full sphere in Eq.~(\ref{SherylCrow}) however.  The left-hand side of Eq.~(\ref{SayHiToYourKnee}) tells us this in a rather deep way: When two points are too far away from each other, at least one of them cannot be in $\cal P$.  We will show this more carefully in the next section:  that the extreme $\|p\drangle\in\cal P$ comprise only part of a sphere.  Of some interest, however, is that Eq.~(\ref{SherylCrow}) already tells us that we cannot have the full sphere as well.  For, the radius of the sphere is such that it exceeds the boundary of the probability simplex $\Delta_{d^2}$.  Hence, at the very least, $\cal P$ is contained within a nontrivial intersection of sphere and simplex.

This is established by a nice argument due to Gabriel Plunk \cite{Plunk02}.  Let us calculate the shortest distance between $\|c\drangle$ and an $n$-flat of the simplex.  The definition of an $n$-flat is that it contains only probability vectors with $n$ vanishing components.  For instance, all $\|p\drangle$ of the form
\be
\|p_n\drangle=\Big(p(1), p(2), \ldots, p(d^2-n), 0, 0, \ldots, 0\Big)^{\!\rm T}\;,
\ee
with $d^2-n$ initial nonvanishing components and $n$ final vanishing components.  A more general $n$-flat would have all the vanishing and nonvanishing components interspersed.

What is the minimal distance $D_{\rm min}\Big(\|c\drangle,\|p_n\drangle\Big)$ between the center point and an $n$-flat?  Taking
\be
D^2\Big(\|c\drangle,\|p_n\drangle\Big)=\sum_{i=1}^{d^2-n}\!\left(p_n(i)-
\frac{1}{d^2}\right)^{\!2}+\sum_{i=d^2-n+1}^{d^2}\!\!\left(0-\frac{1}{d^2}\right)^{\!2}
\ee
generally, and recognizing the constraint
\be
\sum_{i=1}^{d^2-n} p_n(i)=1\;,
\ee
we can use the calculus of variations to find
\be
D_{\rm min}^2\Big(\|c\drangle,\|p_n\drangle\Big)=\frac{n}{d^2(d^2-n)}\;.
\ee
Can there be an $n$ for which
\be
D_{\rm min}^2\Big(\|c\drangle,\|p_n\drangle\Big)< r^2\;?
\ee
In other words, can the sphere ever poke outside of the probability simplex? Just solving the inequality for $n$ gives $n<\frac{1}{2}d(d-1)$.

Thus, when $n<\frac{1}{2}d(d-1)$, the point
\be
\|p_{s(n)}\drangle\equiv\left(\frac{1}{d^2-n},\frac{1}{d^2-n},\ldots,
\frac{1}{d^2-n},0,0,\ldots,0\right) \ee
on an $n$-flat surface of the simplex lies well within the sphere the extreme points of $\cal P$ inhabit.  Only in the case of the qubit, $d=2$, does the sphere reside completely within the simplex---the set is equivalent to the well-known Bloch sphere.

A corollary to Plunk's derivation is that we can put a (weak) bound on the maximum number of zero components a valid $\|p\drangle$ can contain. To have $n$ zero components, $\|p\drangle$ must live on an $n$-flat.  But extreme $\|p\drangle$ are always a distance $D^2_{\rm extreme}=\frac{d-1}{d^2(d+1)}$ from $\|c\drangle$.  So, if $n$ is such that
\be
D^2_{\rm min}\Big(\|c\drangle,\|p_n\drangle\Big)>D^2_{\rm extreme}\;,
\ee
then $\|p\drangle$ can surely not live on the $n$-flat.  This limits $n$:  If $n>\frac{1}{2}d(d-1)$, then a quantum state cannot live on that $n$-flat.
Thus, for a valid $\|p\drangle$, there is an upper bound to how many zero components it can have\footnote{Our first inclination was that this is surely a weak bound.  But even in full-blown quantum mechanics, we know of no better bound than this.  This follows from the best bound we are aware of in that context, a Hilbert-space bound of Delsarte, Goethels, and Seidel \cite{Delsarte75} (which we note can also be proven by elementary Gram matrix methods in Hilbert-Schmidt space).  Let $P_i$, $i=1, ..., v$, be a set of rank-1 projection operators on an $f$-dimensional Hilbert space ${\mathcal H}_f$ such that
$\tr P_i P_j = c$, for all $i\ne j$.
Then
$$
v\le \frac{f(1-c)}{1-fc}\;.
$$
To find the maximum number of zero components $\|p\drangle$ can contain, we just need to ask the question of how many SIC vectors can possibly fit in a $(d-1)$-dimensional subspace.  Inserting the parameters $f=d-1$ and $c=1/(d+1)$ into this bound, we find $n_{\rm zeros}\le \frac{1}{2}d(d-1)$.  Interestingly, this bound is saturated when $d=2$ and $d=3$. On the other hand, in dimensions $d=4$ and $d=5$, D. M. Appleby has checked exhaustively for the known SICs that never more than $d-1$ of the vectors fit within a $(d-1)$-dimensional subspace \cite{Appleby08pc}.}:
\be
n_{\rm zeros}\le \frac{1}{2}d(d-1)\;.
\label{NumZeroes}
\ee
However, an alternative and more direct argument for Eq.~(\ref{NumZeroes}) is this---it is a straightforward application of the Schwarz inequality:
\be
1=\left(\sum_{\stackrel{\rm nonzero}{\rm \scriptscriptstyle terms}}p(i)\right)^{\!2}\le\Big(d^2-n_{\rm zeros}\Big)\!\left(\sum_{\stackrel{\rm nonzero}{\rm \scriptscriptstyle terms}}p(i)^2\right)=\Big(d^2-n_{\rm zeros}\Big)\frac{2}{d(d+1)}\;.
\ee
Eq.~(\ref{NumZeroes}) follows immediately.

But this is only the beginning of the trimming of the Bloch sphere:  More drastic restrictions come from the left-hand inequality of the urungleichung.

\subsection{Consequently an Underlying `Dimensionality'}

\label{MorningOfLittleHope}

What else does the inequality in Eq.~(\ref{SayHiToYourKnee}) imply?  Here is at least one more low hanging fruit. The left side of Eq.~(\ref{SayHiToYourKnee}) signifies that the ``most orthogonal'' two valid distributions $\| p\drangle$ and $\| q\drangle$ can ever be is
\be
\dlangle p\| q\drangle=\sum_i p(i)q(i)=\frac{1}{d(d+1)}\;.
\ee
Their overlap can never approach zero; they can never be truly orthogonal. Now, suppose we have a collection of distributions $\| p_k\drangle$, $k=1,\ldots,n$, all of which live on the sphere---that is, they individually saturate the right-hand side of Eq.~(\ref{SayHiToYourKnee}).  We can ask, how large can the number $n$ can be while maintaining that each of the $\| p_k\drangle$ be maximally orthogonal to each other.  Another way to put it is, what is the maximum number of ``mutually maximally distant'' states?

In other words, we would like to satisfy
\be
\dlangle p_k\| p_l\drangle=\frac{\delta_{kl}+1}{d(d+1)}
\ee
for as many values as possible.  It turns out that there is a nontrivial constraint on how large $n$ can be, and it is none other than $n=d$.

To see this, let us again reference the center of the probability simplex with all our vectors.  Define
\be
\|w_k\drangle=\|p_k\drangle-\|c\drangle\;.
\ee
In these terms, our constraint becomes
\be
\dlangle w_k\| w_l\drangle=\frac{d\delta_{kl}-1}{d^2(d+1)}\;.
\ee
However, notice what this means:  We are asking for a set of vectors whose Gram matrix $G=[\dlangle w_k\| w_l\drangle]$ is an $n\times n$ matrix of the form
\be
G=
\left(
  \begin{array}{cccc}
    a & b & b & b \\
    b & a & b & b \\
    b & b & a & b \\
    b & b & b & a \\
  \end{array}
\right)
\label{113}
\ee
with
\be
a=\frac{d-1}{d^2(d+1)}\qquad\mbox{and}\qquad b=\frac{-1}{d^2(d+1)}\;.
\ee
By an elementary theorem in linear algebra, a {\it proposed\/} set of vectors with a {\it proposed\/} Gram matrix $G$ can exist if and only if $G$ is positive semi-definite \cite[pp.~407--408]{Horn85}.  Moreover, the rank of $G$ represents the number of linearly independent such vectors. (We write ``proposed'' because if $G$ is not positive semi-definite, then of course there are no such vectors.)

Since $G$ in Eq.~(\ref{113}) is a circulant matrix, its eigenvalues can be readily calculated:  one takes the value
\be
\lambda_0=a+(n-1)b=\frac{d-n}{d^2(d+1)}
\ee
while all the $n-1$ others are
\be
\lambda_k=a-b=\frac{1}{d(d+1)}\;.
\ee
To make $G$ positive semi-definite, then, we must have $n\le d$, with $n=d$ being the maximal value.  At that point $G$ is a rank-$(d-1)$ matrix, so that only $d-1$ of the $\| w_l\drangle$ are linearly independent.

On the other hand, all $d$ vectors $\| p_k\drangle=\| w_k\drangle+\| c\drangle$ actually are linearly independent.  To see this, suppose there are numbers $\alpha_i$ such that
$
\sum_i \alpha_i \| p_i\drangle=0\;.
$
Acting from the left on this equation with $\dlangle c\|$, one obtains
\be
\sum_i \alpha_i=0\;.
\ee
On the other hand, acting on it with $\dlangle p_k\|$, we obtain
\bea
0\;=\;\frac{2}{d(d+1)}\alpha_k + \frac{1}{d(d+1)}\sum_{i\ne k}\alpha_i
&=&
\frac{1}{d(d+1)}\alpha_k + \frac{1}{d(d+1)}\sum_i\alpha_i
\nonumber\\
&=& \frac{1}{d(d+1)}\alpha_k\;.
\eea
So indeed,
\be
\sum_i \alpha_i\| p_i\drangle=0\qquad\Longrightarrow\qquad \alpha_k=0 \;\;\forall k\;.
\ee

What this reveals is a significantly smaller ``dimension'' for the valid states on the surface of the sphere than one might have thought. A priori, one might have thought that one could get nearly $d^2$ maximally equidistant points on the sphere, but it is not so---only $d$ instead. This is certainly a suggestive result, but ``dimension'' at this stage must remain in quotes.  Ultimately one must see that the Hausdorff dimension of the manifold of valid extreme states is $2d-2$ (i.e., what it is in quantum theory), and the present result does not get that far.

\subsection{Example: Excising Some Flat Zeros-Bound Vectors}

We saw in Section \ref{Blubbery} that valid $\|p\drangle$ on the surface of the sphere can have no more than $n=\frac{1}{2}d(d-1)$ vanishing components.  Let us ask now whether there are any valid states of the form
\be
\|z\drangle=\frac{2}{d(d+1)}
\Big(\underbrace{0,0,\ldots,0}_{\stackrel{\frac{1}{2}d(d-1)}{\rm \scriptscriptstyle terms}},
\underbrace{1,1,\ldots,1}_{\stackrel{\frac{1}{2}d(d+1)}{\rm \scriptscriptstyle terms}}\Big)^{\rm T}\;,
\ee
where $\|z\drangle$ can also stand for a generic permutation of the vector displayed here.  Let us call these kinds of vectors, ``flat zeros-bound vectors.''

Certainly they cause no trouble with respect to Eq.~(\ref{SayHiToYourKnee}) for the basis states $\|e_k\drangle$.  For, one can check easily that either
\be
\dlangle z\|e_k\drangle=\frac{1}{d(d+1)} \qquad\quad \mbox{or} \qquad\quad
\dlangle z\|e_k\drangle=\frac{1}{d(d+1)}\left(1+\frac{2}{d+1}\right)\;,
\ee
depending on the particular value of $k$ in relation to the $\|z\drangle$ vector.  Neither of these values violate Eq.~(\ref{SayHiToYourKnee})---so they cannot be discounted for that reason.

However the trouble comes when one analyzes the overlaps between the $\|z\drangle$ vectors themselves.  The most extreme case occurs when two vectors $\|z\drangle$ and $\|z^\prime\drangle$ have no zeros at all in corresponding components.  Then $\dlangle z\| z^\prime\drangle$ has only $d$ nonzero terms in the sum and consequently
\be
\dlangle z\| z^\prime\drangle = \frac{1}{d(d+1)}\frac{4}{d+1}\;.
\ee
When $d\ge 4$, $\|z\drangle$ and $\|z^\prime\drangle$ will become too orthogonal for Eq.~(\ref{SayHiToYourKnee}), and so at least one of them must be a invalid state.  More generally, if $\|z\drangle$ and $\|z^\prime\drangle$ match each other in $g$ {\it nonzero} components
\be
\dlangle z\| z^\prime\drangle = g\left(\frac{2}{d(d+1)}\right)^2\;.
\ee
Thus, unless
\be
g\ge \frac{1}{4}d(d+1)\;,
\ee
the two vectors will be too orthogonal to each other for Eq.~(\ref{SayHiToYourKnee}), and at least one of the two must be invalid. In other words, the urungleichung stipulates that at least one of the vectors must be excised from any {\it tentative\/} state-space $\mathcal S$ that might contain the two.

On the other hand, when $d=3$ the argument above carries no force, and there are indeed maximally equidistant flat zeros-bound vectors that correspond to valid quantum states.  Simply look at Eq.~(\ref{Moutard}).  Note for instance that $|\phi_1\rangle=(0,0,1)^{\rm T}$ is orthogonal to all three vectors in the first column and has equal-magnitude overlap with all the remaining elements of the SIC.  Thus $|\phi_1\rangle$ corresponds to one of our $\|z\drangle$ vectors:  $\|z\drangle=\frac{1}{6}(0,1,1,0,1,1,0,1,1)^{\rm T}$.  Similarly, $\frac{1}{6}(1,0,1,1,0,1,1,0,1)^{\rm T}$ and $\frac{1}{6}(1,1,0,1,1,0,1,1,0)^{\rm T}$ also correspond to valid quantum states, as can seen by looking at the second and third columns in Eq.~(\ref{Moutard}).

Yet, even though no two $\|z\drangle$ vectors are too far from each other with respect to the urungleichung, there nevertheless must be other (distinct) considerations leading to the excision of some of the vectors of this form.  For instance, consider $|\psi_1\rangle$, $|\psi_2\rangle$, and $|\psi_3\rangle$ in Eq.~(\ref{Moutard}).  These three vectors are linearly independent, and hence span the whole space. Consequently, there can be no vector of the form $\|p\drangle=(0,0,0,p_4,p_5,p_6,p_7,p_8,p_9)^{\rm T}$ included within $\mathcal S$---which means, in a final analysis, $\|z\drangle=\frac{1}{6}(0,0,0,1,1,1,1,1,1)^{\rm T}$ cannot be included in $\mathcal S$ as well.

\subsection{Summary of the Argument So Far}

\setcounter{assump}{-1}

Let us summarize the assumptions made to this point and summarize their consequences as well.  The paper has become quite unwieldy at this point, and perhaps it is useful to reiterate the main points in a more succinct manner.

\begin{assump}{\rm $\!\!$:}
{\rm The Urgleichung.}  See Figure 2.  Degrees of belief for outcomes in the sky and degrees of belief for outcomes on the ground ought to be related by this fundamental equation:
\be
q(j)=(d+1)\sum_{i=1}^{d^2} p(i) r(j|i) - \frac{1}{d}\sum_{i=1}^{d^2} r(j|i)\;.
\label{Reprise}
\ee
\end{assump}
From the urgleichung, the urungleichung arises by the requirement that in all valid uses of it, $0\le q(j)\le 1$.  The sets $\mathcal P$ and $\mathcal R$ are defined to be sets of priors $\|p\drangle$ and stochastic matrices $R$, that are consistent and maximal.

\begin{assump}{\rm $\!\!$:}
{\rm Principle of Reciprocity:\ Posteriors Are Priors.}  For any $R\in\mathcal R$, a posterior probability consequent upon outcome $j$ of a ground measurement,
\be
\mbox{\rm Prob}(i|j)=\frac{r(j|i)}{\sum_k r(j|k)}
\label{ChickenThigh}
\ee
may be taken as a valid prior $p(i)$ for the outcomes of the measurement in the sky. Moreover, all valid priors $p(i)$ may arise in this way.
\end{assump}

\begin{assump}{\rm $\!\!$:}
{\rm Basis states span the simplex $\Delta_{d^2}$.}  The conditional probabilities $r(j|i)$ derived from setting the ground measurement equal to the sky measurement give rise to posterior distributions Eq.~(\ref{ChickenThigh}) that span the whole probability simplex.
\end{assump}

\begin{assump}{\rm $\!\!$:}
{\rm Extreme-Point Preparations.}
The extreme points of the convex set $\mathcal P$ may all be generated as the posteriors of a suitably chosen ground measurement for which maximal ignorance of sky outcomes implies maximal ignorance of ground outcomes.  Moreover, these measurements all have the minimum number of outcomes consistent with generating the basis distributions $\|e_k\drangle$ in this way.
\end{assump}

With these four assumptions, we derived that the basis distributions $\|e_k\drangle$ should be among the valid states $\cal P$.  We derived that for any $\|p\drangle\in \cal P$, the probabilities are bounded above by $p(k)\le \frac{1}{d}$.  We derived that the extreme points of the valid $\|p\drangle$ should live on the surface of a sphere of interesting radius (a sphere that actually pokes outside the probability simplex). We found a bound on the number of zero components of $\|p\drangle$ that is as good as the best known bound within full-blown quantum mechanics.  And most particularly we derived that for any two valid distributions $\|p\drangle$ and $\|s\drangle$ (including the case where $\|p\drangle=\|s\drangle$), it must hold that
\be
\frac{1}{d(d+1)}\,\le\, \sum_i p(i) s(i) \,\le\, \frac{2}{d(d+1)}\;.
\label{PimpStick}
\ee
From the latter, it follows that no more than $d$ extreme points $\|p\drangle$ can ever be mutually maximally distant from each other. Furthermore we showed that not every flat zeros-bound vector can be a valid $\|p\drangle$.

These are all tasty hints that our structure might just be isomorphic to quantum-state space under the assumption that SICs exist.  But clearly these are just hints.  What really needs to be derived is that the extreme points of such a convex set correspond to an algebraic variety of the form
\be
p(k)=\frac{(d+1)^2}{3d}\sum_{ij} c_{ijk}\,p(i)p(j)-\frac{1}{3d}\;,
\ee
as given in Eq.~(\ref{MegaMorph}), with a set of $c_{ijk}$ obtaining the appropriate properties.  Whether this step can be made without making any further assumptions, we do not know.  Nor do we have a strong feeling presently of whether the auxiliary Assumptions 1, 2, and 3 are the ones best posited for achieving our goal. The key idea, the one to which we would want to hold absolutely fast, however, is to supplement Assumption 0 with as little extra structure as possible for getting all the way to full-blown quantum mechanics.  Clearly much work remains, both at the technical and conceptual level.

\section{Relaxing the Constants and Regaining Them}

\setcounter{resump}{-1}

But what is the origin of the urgleichung in the first place?
In this Section, we take a small step toward a deeper understanding of the particular form our function
\be
q(j)=G\!\left(\sum_i p(i)r(j|i),\sum_i r(j|i)\right)
\ee
takes on in Eq.~(\ref{Reprise}), by initially generalizing away from Eq.~(\ref{Reprise}) and then testing what it takes to get back to it.

What we mean by this is that we should imagine the more general set-up in Figure 2, where the number of outcomes for the measurement in the sky is potentially some more general number $n$ (not initially assumed to be a perfect square $d^2$).  Furthermore we drop away all traces of the parameter $d$, by considering a generalized urgleichung with two initially arbitrary parameters $\alpha$ and $\beta$.  That is to say,
for this Section, our fundamental postulate will be:
\begin{resump}{\rm $\!\!$:} {\rm Generalized Urgleichung}.
For whichever experiment we are talking about for the ground, $q(j)$ should be calculated according to
\be
q(j)=\alpha\sum_{i=1}^n p(i)r(j|i) - \beta \sum_{i=1}^n r(j|i)\;,
\label{BigBoy}
\ee
where $\alpha$ and $\beta$ are fixed nonnegative real numbers.
\end{resump}
Otherwise, all considerations will be the same as they were in the beginning of Section \ref{Heimlich}.  Particularly, the measurements on the ground can have any number $m$ of outcomes, where the value $m$ in any individual case will be set by the details of the measurement under consideration at that time.  Our goal will be to see what assumptions can be added to this basic scenario so that the urgleichung in Eq.~(\ref{Reprise}) re-arises in a natural manner.  That is to say, we would like to see what assumptions can be added to this recipe so that $\alpha=d+1$, $\beta=\frac{1}{d}$, and $n=d^2$ (for some $d$) are in fact part of the end result.

Immediately, one can see that $n$, $\alpha$, and $\beta$ cannot be independent.  This just follows from the requirements that
\be
\sum_{j=1}^m q(j) = 1, \quad \quad\sum_{j=1}^m r(j|i) =1 \ \ \forall i, \quad \mbox{and}\quad \sum_{i=1}^n p(i)=1\;.
\ee
Summing both left and right sides of Eq.~(\ref{BigBoy}) over $j$, one obtains,
\be
n \beta = \alpha-1\;.
\label{RainyDay}
\ee
Furthermore, since $\beta\ne 0$ is assumed, requiring $q(j)\ge 0$ necessitates
\be
\frac{\alpha}{\beta}\ge\frac{\sum_i r(j|i)}{\sum_i p(i)r(j|i)}\ge 1\;.
\ee

As before, we now start studying the consequences of the full requirement that $0\le q(j)\le 1$, in the form of a generalized urungleichung:
\be
0\;\le\; \alpha\sum_{i=1}^n p(i)r(j|i) - \beta \sum_{i=1}^n r(j|i)\;\le\; 1\;.
\label{UrBoy}
\ee
The two sets $\mathcal P$ and $\mathcal R$ are defined analogously to the discussion just after Eq.~(\ref{TheTrueUr}), the first a set of priors for the sky and the second a set of conditionals for the ground (given the outcomes $i$ in the sky).  $\mathcal P$ and $\mathcal R$ are taken to be consistent and maximal.

Two assumptions, we shall borrow straight away from our previous development in Section \ref{Heimlich}.
\begin{resump}{\rm $\!\!$:}
{\rm Principle of Reciprocity:\ Posteriors Are Priors.}  For any $R\in\mathcal R$, a posterior probability, consequent upon outcome $j$ of a ground measurement,
\be
\mbox{\rm Prob}(i|j)=\frac{r(j|i)}{\sum_k r(j|k)}
\label{ChickenButt}
\ee
may be taken as a valid prior $p(i)$ for the outcomes of the measurement in the sky. Moreover, all valid priors $p(i)$ may arise in this way.
\end{resump}
\begin{resump}{\rm $\!\!$:}
{\rm Basis states span the simplex $\Delta_{d^2}$.}  The conditional probabilities $r(j|i)$ derived from setting the ground measurement equal to the sky measurement give rise to posterior distributions $\|e_k\drangle$, via Eq.~(\ref{ChickenButt}), that span the whole probability simplex.
\end{resump}

At this stage, the argument goes just as it did in Section \ref{BengalTiger}.  The only distinction between the results of these two assumptions in the present setting as opposed to the previous is that the basis states are constructed of different constants than previously.  Particularly, the components $e_k(i)$ of these basis states take the form
\be
e_k(i)=\frac{1}{\alpha}(\delta_{ki}+\beta)\;.
\label{MorningBoy}
\ee
It will be noted that in this generalized setting
\be
\sum_i e_k(i)^2=\frac{1}{\alpha^2}\Big(1+2\beta+n\beta^2\Big)\;.
\label{CoffeeBelly}
\ee

Let us now consider afresh the concept of in-step unpredictability for a measurement on the ground with $m$ outcomes ($m\ne n$)---that is, a measurement on the ground such that if one has a flat distribution for the outcomes in the sky, one will also have a flat distribution for the outcomes on the ground.  Let us again denote $r(j|i)$ by $r_{\rm \scriptscriptstyle ISU}(j|i)$ in this special case. Following the manipulations we did before, we must have
\be
\sum_i r_{\rm \scriptscriptstyle ISU}(j|i) = \frac{n}{m}\;.
\ee
By the Principle of Reciprocity, we have that
\be
\mbox{Prob}(i|j)=\frac{m}{n} r_{\rm \scriptscriptstyle ISU}(j|i) \quad\mbox{for}\quad j=1,\ldots,m
\ee
each represent a valid probability distribution for the outcomes of the measurement in the sky.

Suppose now that our prior assignment for the sky so happens to be one of these,
\be
p_k(i)\equiv \mbox{Prob}(i|k)=\frac{m}{n} r_{\rm \scriptscriptstyle ISU}(k|i)\;.
\ee
For the present discussion we will ease our usual convention that outcomes on the ground be labeled by $j$ strictly, allowing now $k$ and $l$ to label ground outcomes as well.
Differing prior assignments---say $p_k(i)$ and $p_l(i)$---will consequently lead to distinct assignments---$q_k(j)$ and $q_l(j)$---for the ground.  I.e., we will have
\be
q_k(j) =  \alpha\sum_i p_k(i) r_{\rm \scriptscriptstyle ISU}(j|i)-\frac{n}{m}\beta
\label{Lambada}
\ee
and similarly coming about for $p_l(i)$.
Now, hold the variable $j$ fixed and not think of it as a dummy index any more. This means that $q_k(j)$, for instance, is just a particular numerical value---at the moment, one should not think of it as a function over the values $j$.  Then it does no harm to make the substitution
\be
r_{\rm \scriptscriptstyle ISU}(j|i)=\frac{n}{m}p_j(i)
\ee
into Eq.~(\ref{Lambada}).  Particularly, with this, one obtains formally that,
\be
q_k(j) = \frac{n}{m}\!\left( \alpha\sum_i p_k(i) p_j(i) -\beta\right).
\label{Lupa}
\ee
In other words, whatever these $\|p_k\drangle$ are, it must be the case that
\be
\frac{\beta}{\alpha}\le\sum_i p_k(i) p_j(i)\le\frac{1}{\alpha}\!\left(\frac{m}{n}+\beta\right).
\label{Remnant}
\ee

Let us now introduce a new notion that we did not make use of in the previous development:  We shall say that a measurement on the ground achieves the {\it ideal of certainty\/} if the urgleichung causes the following to be the case:  That a prior assignment $p_k(i)$ for the sky leads to certainty for one of the outcomes on the ground.  I.e.,
\be
p(i)=p_k(i)\quad\Longrightarrow \quad q_k(j)=\delta_{jk}\;.
\ee

To give an example within standard quantum mechanics, a von Neumann measurement with outcomes $P_j=|j\rangle\langle j|$, satisfying $\langle j|k\rangle=\delta_{jk}$, is such a measurement achieving the ideal of certainty.  For suppose that an agent assigns an initial quantum state $\rho=\frac{1}{d}I$ to the system, that an outcome $j$ was obtained on the ground, and that the measurement in the sky was actually enacted in between.  Then the agent's posterior distribution for the sky will be precisely the SIC representation for $P_j$.  But turn the tables:  If this probability distribution actually were the agent's prior for the sky, though the sky measurement never factualized---i.e., only the measurement on the ground being factualized---then the urgleichung would dictate that the agent should have certainty for outcome $j$ on the ground.

This provides motivation for the following assumption in the more general pre-quantum context:

\begin{resump}{\rm $\!\!$:} {\rm Availability of Certainty.}\footnote{In several axiomatic developments of quantum theory---see for instance \cite{Goyal08} and \cite{Hardy01}---the idea of repeated measurements giving rise to certainty (and the associated idea of ``distinguishable states'') is viewed as fundamental to the whole effort.  The usual justification is that the existence of such kinds of measurement is nearly a self-evident necessity.  However, from the quantum-Bayesian point of view where {\it all\/} measurements are generative of their outcomes---i.e., outcomes never pre-exist the act of measurement---and certainty is always subjective certainty \cite{Caves07}, the consistency of adopting a state of certainty as one's state of belief, even in what is judged to be a repeated experiment, is not self-evident at all.  In fact, from this point of view, why one ever has certainty is the greater of the mysteries.}  There are measurements with in-step unpredictability that achieve the ideal of certainty.
\label{AvailableCertainty}
\end{resump}

For measurements of this type, it follows that
\be
\dlangle p_j\| p_k\drangle=\frac{1}{\alpha}\left(\frac{m}{n}\delta_{jk}+\beta\right)\;.
\label{Delapidate}
\ee
Similar to the considerations of Section \ref{MorningOfLittleHope}, we can ask whether there are any constraints on how large $m$ can be, with respect to $\alpha$, $\beta$, and $n$.  And once again exploring Gram-matrix issues, we will see that indeed there are.

In this case, however, it is convenient to first re-express our vectors $\| p_k\drangle$ as expansions with respect to the basis states $\| e_i\drangle$ in Eq.~(\ref{MorningBoy}).
%\footnote{{\tt Challenge for R\"udiger:} Explain in this footnote why %we actually need to make this substitution.  Why couldn't we have just %looked at the Gram matrix formed from the vectors %$\|w_k\drangle=\|p_k\drangle-\|c\drangle$, where $\|c\drangle$ is the %flat probability over $n$ outcomes, as we did in Section 5.4.  Why our %present method works is clear enough to me.  But why the more obvious %method does not, I still find a bit mysterious.}
Write
\be
\| p_k\drangle=\sum_i \alpha_k(i) \| e_i\drangle\;.
\ee
We can do this because the $\| e_i\drangle$ span the simplex.  Of course in this representation the $\alpha_k(i)$ need not be nonnegative numbers, but using Eq.~(\ref{RainyDay}), one can see it is still the case that
\be
\sum_i \alpha_k(i)=1\;.
\ee
Nearly trivially too,
\be
\alpha_k(i)=\alpha\, p_k(i) - \beta\;.
\ee
Transforming $p_k(i)$ to the new variable $\alpha_k(i)$ in  Eq.~(\ref{Delapidate}), we obtain this requirement for a measurement achieving the ideal of certainty:
\be
\sum_i \alpha_j(i)\alpha_k(i)=\alpha\frac{m}{n}\delta_{jk}-\beta\;.
\label{ToTheEastOffice}
\ee
Once again, it is that a set of inner products achieve a certain set of values.

Let us suppose that $m$ is the largest value for which this can be true, at the same time as not violating the requirement that the $\|p_k\drangle$ satisfy the constraints of being probability distributions.  As we will see, $m+1$ vectors can actually be made to satisfy Eq.~(\ref{ToTheEastOffice}), but when so, one of the vectors is necessarily {\it not\/} a probability distribution.

Thus let us ask for the existence of an $(m+1)\times(m+1)$ Gram matrix
of the form
\be
\left(
  \begin{array}{cccc}
    a & b & b & b \\
    b & a & b & b \\
    b & b & a & b \\
    b & b & b & a \\
  \end{array}
\right)
\ee
with diagonal entries $a=\frac{m}{n}\alpha-\beta$ and off-diagonal entries $b=-\beta$.  At what value of $m$ will this matrix stop being positive semi-definite?  Again we use the formula for the eigenvalues of a circulant matrix to answer this question: One eigenvalue takes the value
\be
\lambda_0=a+mb=\alpha\frac{m}{n}-(m+1)\beta
\ee
while all the other $m$ eigenvalues are
\be
\lambda_k=a-b=\alpha\frac{m}{n}\;.
\ee
Setting $\lambda_0=0$ gives then the following relation between $\alpha$, $\beta$, and $n$ and this extremal $m$:
\be
\frac{\alpha}{\beta}=\frac{n(m+1)}{m}\;.
\label{BleedinHorse}
\ee

With this, let us return to the claim that at most $m$ vectors satisfying the inner products Eq.~(\ref{Delapidate}) can actually correspond to probability distributions.  For, suppose we did have $m+1$ distributions $\|p_k\drangle$ satisfying Eq.~(\ref{Delapidate}).  Since they are linearly {\it dependent}, one must be of the form:
\be
\|p\drangle=\sum_{k=1}^{m} \tau_k \|p_k\drangle\;.
\label{Humph}
\ee
For $\|p\drangle$ to be a probability distribution at all (i.e., have terms that sum to 1), we must have $\sum_k \tau_k=1$.  But then, inserting (\ref{Humph}) into (\ref{Delapidate}) for each $l=1,\ldots,m$, gives that
\be
-\beta=\frac{m}{n}\alpha\,\tau_l - \beta\sum_k \tau_k,
\ee
requiring that $\tau_l=0$.  That is, assuming Eqs.~(\ref{Delapidate}) and (\ref{Humph}), along with the requirement that the $p(i)$ sum to 1, leads to a contradiction.  Hence we have for measurements satisfying in-step unpredictability at the same time as the ideal of certainty, the largest value $m$ which can obtain is constrained by Eq.~(\ref{BleedinHorse}).

With this, we are ready to make our final assumption.
Suppose we consider a measurement satisfying Resumption \ref{AvailableCertainty}, and suppose it has a number of values $m$ that is the maximum that quantity can be, i.e., $m$ satisfies Eq.~(\ref{BleedinHorse}).  Call that class ${\mathcal R}_c$.

\begin{resump}{\rm $\!\!$:} {\rm Principle of Identification.}
The class ${\mathcal R}_c$ is nonempty and in it there is an ISU measurement for which {\it one\/} of its posteriors $\mbox{\rm Prob}(i|j_0)=\frac{m}{n} r_{\rm \scriptscriptstyle ISU}(j_0|i)$ has the form of a basis distribution in Eq.~(\ref{MorningBoy}).
\end{resump}

With this assumption in place, there must be a consistency between the $j=k$ case of Eq.~(\ref{Delapidate}) and the previously calculated purity of the basis states Eq.~(\ref{CoffeeBelly}).  Working that out, one obtains a new, independent relation for our variables:
\be
\frac{m}{n}\alpha-\beta=1\;.
\label{NimbleNooph}
\ee

As an aside, notice the following.  If, contrary to the spirit of this section, $\beta=0$ had been assumed at the outset, Eq.~(\ref{RainyDay}) would have dictated that $\alpha=1$.  But then Eq.~(\ref{NimbleNooph}) would have given in turn that $n=m$.  In the language of Hardy's ``Quantum Theory From Five Reasonable Axioms,'' paper \cite{Hardy01}, the number of ``degrees of freedom'' would be equal to the number of ``distinguishable states.''  But, as he notes in the development of his axiom system, quantum theory has a more interesting relation than that.\footnote{In the Hardy development, the best one can say without invoking the ``axiom of simplicity'' is that $n=m^r$ for some integer $r$.  Hardy classifies $r=1$ a ``classical'' theory and $r=2$ a ``quantum theory.''  The remainder of the hierarchy he considers an interesting venue to explore.} We move to that final demonstration presently.

\subsection{Putting the Ingredients Together}

In total, here is what we have.  The consistency of the generalized urgleichung, Resumption 0, requires
\be
n \beta = \alpha-1\;.
\ee
The assumption of a maximal-cardinality measurement satisfying Resumption 3, the availability of certainty, along with the principle of reciprocity, Resumption 1, gives
\be
\alpha m = n(m+1)\beta\;.
\ee
Finally, assuming the principle of identification, Resumption 4, along with the particular form of the basis states given by Resumption 2, gives
\be
\alpha m = n + n\beta\;.
\ee
Simultaneously solving these three equations (and making the substitution $m\rightarrow d$ in order to emphasize the connection back to our quantum mechanical target), we get
\be
n = d^2 \qquad\mbox{and}\qquad \beta = \frac{1}{d}
\qquad\mbox{and}\qquad
\alpha = d + 1\;.
\ee
The standard urgleichung of Eq.~(\ref{Reprise}) is regained.

Let us reiterate slightly the philosophy here.  The generalized urgleichung is assumed universal for all measurements.  What we have done then is considered the assumption of a single very specialized measurement to fix $\alpha$ and $\beta$ in terms of $m$.  Furthermore, we have shown that this special measurement can only exist if $n$ is a perfect square. That we could pick up such a distinction between $n$ and $m$ through these very simple considerations (thinking again of the Hardy discussion at the end of the last subsection), we find quite remarkable.

\section{Summary: From Quantum Interference to Bayesian Coherence}

In this paper, we hope to have given a new and useful way to think of quantum interference:  Particularly, we have shown how to view it as an addition to Bayesian coherence, operative when one calculates probabilities for the outcomes of a factualizable quantum experiment in terms of one explicitly assumed counterfactual.  We did this and never once did we use the idea of a probability {\it amplitude}.  Thus we believe we have brought the idea of quantum interference formally much closer to its root in probabilistic considerations than had been done previously.  For this, we were certainly aided by the elegance of the mathematical machinery of SIC measurements.

In any case, in so doing we made it clear that the Born Rule can be viewed as a relation between probabilities, rather than the usual thing one thinks it is---that it is a {\it setter\/} of probabilities from something more firm or secure than probability itself.  That is, that the Born Rule facilitates a probability assignment from {\it the\/} quantum state.  From the quantum-Bayesian point of view there is no such thing as {\it the\/} quantum state, there being as many quantum states for a system as there are agents interested in considering it.  This last point makes it particularly clear why we needed a way of viewing the Born Rule as an extension of coherence:  One can easily invent situations where two agents will update to divergent quantum states (even pure states, and even orthogonal pure states at that!)\ by looking at the same empirical data \cite{Fuchs02,Fuchs04,Fuchs09}---a quantum state is always ultimately dependent on the agent's priors.

So, we believe we have had some success.  But there is much more to do still.  We gave an indication that the urgleichung and considerations to do with it already specifies a significant fraction of the structure of quantum states---and for that reason one might want to take it as one of the most fundamental axioms of quantum mechanics.  We did not, however, get all the way back to a set based on the manifold of pure quantum states, Eq.~(\ref{MegaMorph}).  Can it be done through more careful thinking without adding any question-begging assumptions?  We do not know.  Moreover, what calls for the urgleichung in the first place?  It is simple, but it is still a strange looking equation.  It would be wonderful, for instance, if one could justify it in terms of bought and returned lottery tickets consequent upon the nullification step in our standard scenario.  Then the positive content of the Born Rule might be viewed as a kind of cost excised whenever one factualizes a SIC.  But this is just speculation.

What is firm is that we have a new setting for quantifying the old idea that, in quantum mechanics, unperformed measurements have no outcomes.

\section{Forward to an Eventual Foreword!}

\begin{flushright}
\baselineskip=13pt
\parbox{2.8in}{\baselineskip=13pt\small
Of every would be describer of the universe one has a right to ask immediately two general questions.  The first is: ``What are the materials of your universe's composition?'' And the second: ``In what manner or manners do you represent them to be connected?''}
\medskip\\
\small --- William James, 1903--1904, \cite{James88}
\end{flushright}

This paper has focussed on adding a new girder to the developing structure of quantum Bayesianism (`QBism' hereafter).  As such, we have taken much of the previously developed program as a background for the present efforts.  If a reader wants to know, for instance, the core arguments for why we choose a more `personalist Bayesianism' rather than a so-called `objective Bayesianism,' she should refer to Refs.~\cite{Fuchs02,Fuchs04,Struggles}.  If she wants to know why a subjective, personalist account of certainty is crucial for breaking the impasse set by the EPR criterion of reality, she should refer to Refs.~\cite{Caves07,Struggles}.  Similarly for other questions on the program.  This paper was mostly devoted to getting a new phase of QBism off the ground without dwelling too much on the past.

Still, fearing James' injunction, we know it is our duty to discuss anew a term in the present paper that we have been waiving about most uncritically:  It is `measurement.'  How can one really understand the proclamation `Unperformed measurements have no outcomes!'\ in a deep, soul-satisfying way?  Answering this question, we feel, is the first step toward characterizing ``the materials of our universe's composition.''

Believe it or not, we take our cue from John Bell:  For, despite our liberal use of the term so far, we think the word `measurement' should indeed be banished from fundamental discussions of quantum theory~\cite{Bell90}.  However, it is not at all because the word is ``unprofessionally vague and ambiguous,'' as Bell said of it~\cite{Bell87}.\footnote{To be sure, the are plenty of things vague and inconsistent in the writings of Bohr, Pauli, Heisenberg, von Weizs\"acker, Peierls, and Peres (some of the most well-known representatives of this so-said `orthodoxy'); we would be the last to take issue with this.  But this does negate our own, particular reasons for disinclination to the word `measurement.' Appleby \cite{Appleby05a} makes the point very well.}  Rather, it is because the word's usage engenders a deep misunderstanding of the very subject matter of quantum theory.\footnote{Note that we were careful to say that measurement ``{\it should be\/} banished.'' For we harbor no illusions that the word will actually disappear from common usage any time soon, even our own, and even our own in the remainder of this very paper!  The point is a rhetorical one, of course, with the aim of getting the reader to rethink the meaning of quantum measurement.  For an argument in some sympathy with our own, but with a different rhetorical flourish, see N.~D. Mermin's ``In Praise of Measurement'' \cite{Mermin06}.}  Quantum theory is a smaller theory than one might think---it is smaller precisely because it indicates the world to be a bigger, more varied place than much of the present philosophy of science can imagine.

To make the point dramatic, let us put quantum theory to the side for a moment, and consider instead basic Bayesian probability theory.  There the subject matter is an agent's expectations for this and that.  For instance, an agent might write down a joint probability distribution $P(h_i,d_j)$ for various mutually exclusive hypotheses $h_i$, $i=1,\ldots,n$, and data values $d_j$, $j=1,\ldots,m$, appropriate to some phenomenon.  As we have already discussed at great length, a major role of the theory is that it provides a scheme (coherence) for how these probabilities should be related to other probabilities, $P(h_i)$ and $P(d_j)$ say, as well as to any other degrees of belief the agent has for other phenomena.  The theory also prescribes that if the agent is given a specific data value $d_j$, he should update his expectations for everything else within his interest.  For instance, under the right conditions \cite{Skyrms87b}, he should reassess his probabilities for the $h_i$ by conditionalizing:
\be
P_{\mbox{new}}(h_i)=\frac{P(h_i,d_j)}{P(d_j)}\;.
\ee
But what is this phrase ``given a specific data value''?  What does it really mean in detail?  Shouldn't one specify a mechanism or at least a chain of logical and/or physical connectives for how the raw fact signified by $d_j$ comes into the field of the agent's consciousness?  And who is this ``agent'' reassessing his probabilities anyway?  Indeed, what is the precise definition of an agent? How would one know one when one sees one?  Can a dog be an agent? Or must it be a person? Maybe it should be a person with a PhD?\footnote{Tongue-in-cheek reference to Bell again \cite{Bell90}.}

With a touch of sarcasm, we are led to ask:  Shouldn't probability theory be held accountable for giving answers to all these questions?  In other words, shouldn't a book like L.~J. Savage's {\sl The Foundations of Statistics\/} \cite{Savage54} spend some of its pages demonstrating how the axioms of probability---by way of their own power---give rise, at least in principle, to agents and data acquisition itself?  Otherwise, shouldn't probability theory be charged with being ``unprofessionally vague and ambiguous''?

Well, that's a stupid expectation, and anyone who reads this can sense it.  Probability theory has no chance of answering these questions because they are not questions within the subject matter of the theory.  Within probability theory, the notions of ``agent'' and ``given a data value'' are primitive and irreducible and have no shame for being so. Guiding agents' decisions based on data is what the whole theory is constructed for---just like primitive forces and primitive masses are what the whole theory of classical mechanics is constructed for.  As such, agents and data are the most exalted elements within the structure of probability theory---they are not to be constructed from it, but rather the former are there to receive the theory's guidance, and the latter are there to designate that the world external to the agent has not, after all, disappeared in a dream.

So ends a bit of a diatribe.  We present this heavy rhetoric because there is an important lesson to be learned from it, and sometimes one needs to be hit over the head for something new to take hold in one's worldview.  QBism says if all of this is true of Bayesian probability theory in general, it is true of quantum theory as well.  As the foundations of probability theory dismisses the questions of where data comes from and what constitutes an agent, so can the foundations of quantum theory dismiss them too.  This last point is one of the strong reasons for making the QBist move in the first place.

But so said, there will surely be an outcry in the majority of readers at this point.  For, a rebuttal going something like this will be formulated immediately: ``It is one thing to say all this of probability theory, but quantum theory is a wholly different story.''  And a sound will go out to all the lands, ``Quantum mechanics is no simple branch of mathematics, be it probability or statistics.  Nor can it plausibly be a theory about the insignificant specks of life in our vast universe making gambles and decisions. Quantum mechanics is one of our best theories of the world!  It is one of the best maps we have drawn yet of what is actually out there.''\footnote{This is hardly conclusive evidence, but if one does a Google search on the simultaneous terms ``quantum mechanics'' and ``best theories of the world'' one will find about 22,000 hits.  If one does a search on the terms ``quantum'' and ``best theories of the world'' one will find about 36,000 hits.}  But this is just where the average reader errs, and that is all it reveals, we say.  We hold fast:  Quantum theory is simply {\it not\/} a `theory of the world.'  Just like probability theory is not a theory of the world, quantum theory is not as well:  It is a theory for the use of agents immersed in and interacting with a world of a particular character, the quantum world.

The details in the last line are crucial for understanding what we are trying to say; we hope the reader will ponder over them for a moment.  We certainly have no bone to pick with the idea of a world external to the agent.  Indeed it must be as Martin Gardner says \cite{Gardner83},
\bq
The hypothesis that there is an external world, not dependent on
human minds, made of {\em something}, is so obviously useful and
so strongly confirmed by experience down through the ages that we
can say without exaggerating that it is better confirmed than any
other empirical hypothesis.  So useful is the posit that it is almost impossible for anyone except a madman or a professional metaphysician to comprehend a reason for doubting it.
\eq
Yet there is no implication in these words that quantum theory, for all its success in chemistry, physical astronomy, laser making, and so much else, must be read off as a theory of the world.  There is room for a significantly more interesting form of dependence:  It can be (and is) that quantum theory is conditioned by the character of the world, but yet is not a theory directly of it.  Confusion on this very point, we believe, is what has caused most of the discomfort in quantum foundations in the 82 years since the theory's coming to a relatively stable equilibrium in 1927.

Let us now expand on this conception of quantum mechanics by going back to our discussion of Bell and the word ``measurement.''  As we have already made amply clear, it is hardly because we think it unprofessionally vague and ambiguous that we wish ``measurement'' banished.  Instead we wish it because the word subliminally whispers the philosophy of its birth:  That quantum mechanics {\it should\/} be conceived in a way that makes no ultimate reference to agency, and that agents become constructed out of the theory, rather than taken as the primitive entities the theory is meant to aid.  In a nutshell, the word deviously carries forward the impression that quantum mechanics should be viewed as a theory directly of the world.

One gets a sense of the boundaries the word ``to measure'' places upon our interpretive thoughts by turning to any English dictionary.  Here is a sampling from {\tt http://dictionary.com/}:
\begin{verse}
$\bullet$
to ascertain the extent, dimensions, quantity, capacity, etc., of,
esp.\ by comparison with a standard
\\
$\bullet$ to estimate the relative amount, value, etc., of, by comparison with some standard
\\
$\bullet$ to judge or appraise by comparison with something or someone else
\\
$\bullet$
to bring into comparison or competition
\end{verse}
In not one of these definitions do we get an image of anything being created in the measuring process; none give any inkling of the crucial contextuality of quantum measurements, the context being a parameter ultimately set only in terms of the agent.  Measurement, in its common usage, is something passive and static:  it is comparison between {\it existents}.  No wonder a slogan like `Unperformed measurements have no outcomes!'\ would seem irreparably paradoxical.  If a quantum measurement is not comparison, but something else, the only way out of the impasse is to understand what that something else is.

Fixing this word is the prerequisite to a new ontology---in other words, prerequisite to a statement about the (hypothesized) character of the world that does not make direct reference to our actions and gambles within it.  Therefore, as a start, let us rebuild quantum mechanics in terms more palatable to a quantum Bayesian.

\subsection{The Paulian Idea and the Jamesian Multiverse}

\medskip

\begin{flushright}
\baselineskip=13pt
\parbox{4.2in}{\baselineskip=13pt\small
The value of a multiverse,$^{31}$ as compared with a universe, lies in this, that where there are cross-currents and warring forces our own strength and will may count and help decide the issue; it is a world where nothing is irrevocably settled, and all action matters. A monistic world is for us a dead world; in such a universe we carry out, willy-nilly, the parts assigned to us by an omnipotent deity or a primeval nebula; and not all our tears can wipe out one word of the eternal script. In a finished universe individuality is a delusion; ``in reality,'' the monist assures us, we are all bits of one mosaic substance. But in an unfinished world we can write some lines of the parts we play, and our choices mould in some measure the future in which we have to live. In such a world we can be free; it is a world of chance, and not of fate; everything is ``not quite''; and what we are or do may alter everything.}
\medskip\\
\small --- Will Durant on William James's `pluralism' \cite{Durant53}
\end{flushright}

\footnotetext{Beware: This usage of multiverse is not remotely similar to the Everettian use of the same, which, from our point of view, is quite a bastardization of William James's original multiverse.  (The {\sl Oxford English Dictionary\/} credits James with originating the word in 1895.)  The Everettian multiverse \cite{Deutsch97,Wallace02,Greaves04}, with its ``wave function of the universe'' and ``complete determinism,'' is about about as pure an example of a monism as one can ever hope to see.}

\addtocounter{footnote}{1}

\begin{figure} %\leavevmode
\begin{center}
\includegraphics[height=3.75in]{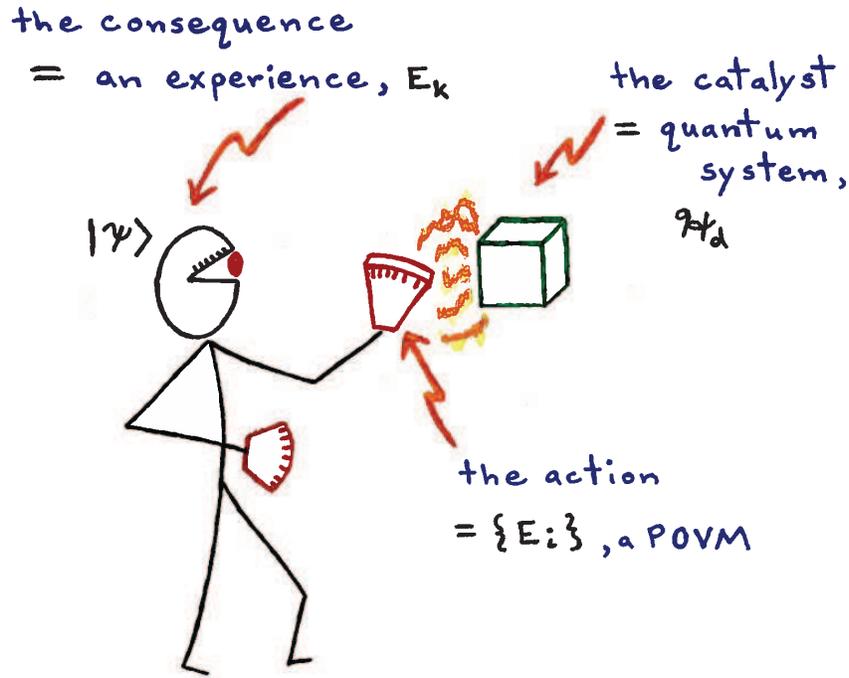}
%paulian2.eps}
\bigskip\caption{The Paulian Idea \cite{Fuchs01}---in the form of a figure inspired by John Archibald Wheeler \cite{WheelerFig}.  In contemplating a quantum measurement (though the word is a misnomer), one makes a conceptual split in the world:  one part is treated as an agent, and the other as a kind of reagent or catalyst (one that brings about change in the agent itself).  In older terms, the former is an observer and the latter a quantum system of some finite dimension $d$.  A quantum measurement consists first in the agent taking an {\it action\/} on the quantum system.  The action is formally captured by some POVM $\{E_i\}$. The action leads generally to an incompletely predictable {\it consequence}, a particular personal experience $E_i$ for the agent \cite{FuchsDelirium}.  The quantum state $|\psi\rangle$ makes no appearance but in the agent's head; for it  only captures his degrees of belief concerning the consequences of his actions, and---in contrast to the quantum system itself---has no existence in the external world.  Measurement devices are depicted as prosthetic hands to make it clear that they should be considered an integral part of the agent.  (This contrasts with Bohr's view where the measurement device is always treated as a classically describable system external to the observer.)  The sparks between the measurement-device hand and the quantum system represent the idea that the consequence of each quantum measurement is a unique creation within the previously existing universe \cite{FuchsDelirium}.  Two points are decisive in distinguishing this picture of quantum measurement from a kind of solipsism:  1) The very conceptual split of agent and external quantum system: If it were not needed, it would not have been made. Imagining a quantum measurement without a quantum system participating would be as paradoxical as the Zen koan of the sound of one hand clapping. 2) Once the agent chooses an action $\{E_i\}$ to take, the particular consequence $E_k$ of it is beyond his control---that is, the actual outcome is not a product of his whim and fancy.  Wolfgang Pauli characterized this picture as a ``wider form of the reality concept'' than that of Einstein's, which he labeled ``the ideal of the detached observer'' \cite{Pauli94,Laurikainen88,Gieser05}. What is important for modern developments is that the particular character of the catalysts---i.e., James's ``materials of your universe's composition''---must leave its trace in the formal rules that allow us to conceptualize factualizable measurements in terms of a standard counterfactual one.  The explorations in this paper represent a first step in that direction.}
\end{center}
\end{figure}

The best way to begin a more thoroughly QBist delineation of quantum mechanics is to start with two choice quotes on personalist Bayesianism itself.  The first is from Hampton, Moore, and Thomas \cite{Hampton73},
\bq
Bruno de Finetti believes there is no need to assume that the probability of some event has a uniquely determinable value.  His philosophical view of probability is that it expresses the feeling of an individual and cannot have meaning except in relation to him.
\eq
and the second from D.~V. Lindley \cite{Lindley82},
\bq
The Bayesian, subjectivist, or coherent, paradigm is egocentric.  It is a tale of one person contemplating the world and not wishing to be stupid (technically, incoherent).  He realizes that to do this his statements of uncertainty must be probabilistic.
\eq
These two quotes make it absolutely clear that personalist Bayesianism is a ``single-user theory.''  Thus, QBism must inherit at least this much egocentrism in its view of quantum states $\rho$.  We hope we made this quite clear in the previous development of the paper.  But here we need to cover new ground:  For, the ``Paulian Idea'' \cite{Fuchs01}---which is also essential to the QBist view---goes further still.  It says that the outcomes to quantum measurements are single-user as well!  That is to say, when an agent writes down her degrees of belief for the outcomes of a quantum measurement, what she is writing down are her degrees of belief about her potential {\it personal\/} experiences arising in consequence of her actions upon the external world \cite{FuchsDelirium}.

Before exploring this further, let us partially formalize in a quick outline the structure of quantum mechanics from this point of view, at the moment retaining the usual mathematical formulation of the theory, but starting the process of changing the English descriptions of what the term ``quantum measurement'' means. \vspace{-.2in}
\bq
\begin{enumerate}
\item
Primitive notions: a) the agent, b) things external to the agent, or, more commonly, ``systems,'' c) the agent's actions on the systems, and d) the consequences of those actions for her experience.

\item
The formal structure of quantum mechanics is a theory of how the agent ought to organize her Bayesian probabilities for the consequences of all her potential actions on the things around her.  Implicit in this is a theory of the structure of actions.  The theory works as follows:

\item
When the agent posits a system, she posits a Hilbert space ${\mathcal H}_d$ as the arena for all her considerations.

\item
Actions upon the system are captured by positive-operator valued measures $\{E_i\}$ on ${\mathcal H}_d$.  Potential consequences of the action are labeled by the individual elements $E_i$ within the set.\footnote{There is a formal similarity between this and the development in Cox \cite{Cox61}, where ``questions'' are treated as sets, and ``answers'' are treated as elements within the sets.}  I.e.,
$$
\mbox{ACTION}=\{E_i\} \qquad \mbox{and} \qquad \mbox{CONSEQUENCE}= E_k\;.
$$

\item
Quantum mechanics organizes the agent's beliefs by saying that she should strive to find a single density operator $\rho$ such that her degrees of belief will always satisfy
$$
\mbox{Prob}\Big(\mbox{CONSEQUENCE}\,\Big|\,\mbox{ACTION}\Big)=
\mbox{Prob}\Big(E_k\,\Big|\,\{E_i\}\Big)=\tr\rho E_k\,,
$$
no matter what action $\{E_i\}$ is under consideration.

\item
Unitary time evolution and more general quantum operations (completely positive maps) do not represent objective underlying dynamics, but rather address the agent's belief changes accompanying the flow of time, as well as belief changes consequent upon any actions taken.

\item
When the agent posits {\it two\/} things external to herself, the arena for all her considerations becomes ${\mathcal H}_{d_1}\otimes{\mathcal H}_{d_2}$.  Actions and consequences now become POVMs on ${\mathcal H}_{d_1}\otimes{\mathcal H}_{d_2}$.

\item
The agent can nonetheless isolate the notion of an action on a single one of the things alone:  These are POVMs of the from $\{E_i\otimes I\}$, and similarly with the systems reversed $\{I \otimes E_i\}$.

\item
Resolving the consequence of an action on {\it one\/} of the things may cause the agent to update her expectations for the consequences of any further actions she might take on the {\it other\/} thing.  But for those latter consequences to come about, she must elicit them through an actual action on the second system.
\end{enumerate}
\eq

The present paper, of course, has predominantly focussed on Item 5 in this list, rewriting the point in purely probabilistic terms---this is what all the previous Sections were about, though we unreservedly used the standard language of ``measurement'' in that part of the development.  With regard to the discussion in the present Section, however, the main points to note are Items 4, 7, 8, and 9.  The common usage of the word ``measurement'' goes out the door precisely in this:  One should not think of the quantum dispensation of it as a comparison, but simply as an {\it action\/} upon the system of interest.  Actions lead to consequences within the experience of the agent, and that is what a quantum measurement is.  A quantum measurement finds nothing, compares nothing, but very much {\it makes\/} something.

It is an absolutely simple linguistic move, but it does nearly infinite work for resetting the quantum-foundations debate.  Particularly here:
For it might indeed have been the case that all this nonstandard formulation was for nought, turning out to be superfluous.  That is, though we have spelled out very carefully in Item 9 that, ``for those latter consequences to come about, she must elicit them through an actual action on the second system,'' maybe there would have been nothing wrong in thinking of the latter (and by analogy the former) quantum measurement as finding a pre-existing value after all.  But this, we have argued previously \cite{Caves07,Struggles} would contradict Item 8, i.e., that one can isolate a notion of an action on a single system alone.  It is in essence the Stairs \cite{Stairs83} variant of the Kochen-Specker argument (with the aid of entanglement) that blocks this interpretation.

Thus, in a QBist painting of quantum mechanics, quantum measurements are ``generative'' in a very real sense.  But by that turn, the consequences or our actions on physical systems must be quite personal and egocentric as well.  Measurement outcomes come about for the agent himself.  Anyone else want an outcome?  Then they must go out and elicit {\it those\/} outcomes themselves through their own actions.  It is in this sense that quantum mechanics is a single-user theory through and through---first in the usual Bayesian sense with regard to personal beliefs, but second in that quantum measurement outcomes are wholly personal experiences.\footnote{The usual belief otherwise---for instance in Pauli's own formulation (which is ultimately inconsistent with his taking measurement devices to be like prosthetic hands), that ``the objective character of the description of nature given by quantum mechanics [is] adequately guaranteed by the circumstance that \ldots\ the results of observation, {\it which can be checked by anyone}, cannot be influenced by the observer, once he has chosen his experimental arrangement'' [italics ours, for pinpointing the offending portion of the formulation]---we state for completeness, is the ultimate source of the Wigner's friend paradox.  This will be expanded upon in a later work by the authors; for the moment see \cite{Struggles}.}

``Rubbish!,'' we can already hear an Australian friend say!  ``It would mean quantum mechanics collapses into a kind of solipsism---a theory that there is only the self.''  But it is not so at all.  By saying quantum mechanics is a single-user theory, we only mean, ``I can use it, you can use it, she can use it---any of us.''  And when any of us uses it, we are using it to better prepare for our own (personal) encounters with the world.  We use it to make sure our {\it beliefs\/} about the consequences of our encounters are {\it consistent\/} with each other.  We mean it is a calculus for ``consistifying'' each of our beliefs for what we think will happen to us.  There is no implication in this whatsoever that there is only one self.

Particularly with regard to the Paulian Idea there are two points that are decisive for dismissing the charge of solipsism \cite{Fuchs02b}.  One is the conceptual split of the world into two parts---one an agent and the other an external quantum system---that gets the discussion of quantum measurement off the ground in the first place.  If such a split were not needed for making sense of the question of actions (actions upon what?\ in what?\ with respect to what?), it simply would not have been made. Imagining a quantum measurement without an autonomous quantum system participating in the process would be as paradoxical as the Zen koan of the sound of a single hand clapping.  The second point is that once the agent chooses an action $\{E_i\}$, the particular consequence $E_k$ of it is beyond his control.  That is to say, the particular outcome of a quantum measurement is not a product of his desires, whims, or fancies---this is the very reason he uses the calculus of probabilities in the first place: they quantify his uncertainty \cite{Lindley06}, an uncertainty that, try as he might, he cannot get around.  So, implicit in this whole picture---this whole Paulian Idea---is an ``external world \ldots\ made of {\it something},'' just as Martin Gardner calls for.  It is only that quantum theory is a rather small theory:  Its boundaries are set by being a handbook for agents immersed within that ``world made of {\it something}.''\footnote{Of course, a gut reaction of many will be to say, ``Well then, it is incomplete after all.  Go seek hidden variables!''  But that is to misunderstand the problematic here.  Theories of decision that really are theories of decision just don't ``port'' to theories or visions of the world in that way.  From the point of view of being a theory for taking actions and gambles, quantum theory is already all that it can be.}

But a small theory can still have grand import, and quantum mechanics most certainly does.  This is because it at least tells us how an ``insider'' sees his role in the world.  Even if quantum mechanics---viewed as an addition to probability theory---is not a theory of the world itself, it is certainly conditioned by the particular character of this world.  Just look at the simplest case of the urgleichung:
$$
q(j)=(d+1)\sum_{i=1}^{d^2} p(i) r(j|i) - 1\;.
$$
It is not one of the infinity of other relations contemplated in the discussion in Section \ref{OhBoy}:
$$
q(j)=G\Big(\{p(i)\},\{r(j|i)\}\Big)\;.
$$
To that extent, even though quantum theory is now understood as a theory of acts, decisions, and consequences \cite{Savage54}, it still possesses essential empirical content.  In code, it tells us about the character of our particular world.  Apparently, the world is made of a stuff that does not have ``consequences'' waiting around to fulfill our ``actions''---it is a world in which the consequences are generated on the fly.  When we on the inside prod that stuff on the outside, the world comes to something that neither side could have foretold.  If our hunch about taking the urgleichung as the main postulate of quantum mechanics is on the right track, then we start to get a sense of how the grand bastions of ontology can still be found in something so tiny as a theory of good personal decisions.

Indeed, one starts to get a sense of a world picture that is part personal---{\it truly\/} personal---and part the joint product of all that interacts.  It is almost as if one can hear in the very formulation of the Born Rule one of William James's many lectures on chance and indeterminism.  Here is one example \cite{JamesDilemma},
\bq
[Chance] is a purely negative and relative term, giving us no information about that of which it is predicated, except that it
happens to be disconnected with something else---not controlled,
secured, or necessitated by other things in advance of its own actual
presence. \ldots\ What I say is that it tells us
nothing about what a thing may be in itself to call it ``chance.'' \ldots\ All you mean by calling it ``chance'' is that this is not guaranteed, that it may also fall out otherwise. For the system of other things has no positive hold on the chance-thing. Its origin is in a certain fashion negative: it escapes, and says, Hands off!\ coming, when it comes, as a free gift, or not at all.

This negativeness, however, and this opacity of the chance-thing when thus considered {\it ab extra}, or from the point of view of previous things or distant things, do not preclude its having any amount of positiveness and luminosity from within, and at its own place and moment. All that its chance-character asserts about it is that there is something in it really of its own, something that is not the unconditional property of the whole. If the whole wants this
property, the whole must wait till it can get it, if it be a matter
of chance. That the universe may actually be a sort of joint-stock
society of this sort, in which the sharers have both limited
liabilities and limited powers, is of course a simple and conceivable
notion.
\eq
And here is another \cite{JamesHegelisms},
\bq
Why may not the world be a sort of republican banquet of this sort, where all the qualities of being respect one another's personal sacredness, yet sit at the common table of space and time?

To me this view seems deeply probable.  Things cohere, but the act of cohesion itself implies but few conditions, and leaves the rest of their qualifications indeterminate.  As the first three notes of a tune comport many endings, all melodious, but the tune is not named till a
particular ending has actually come,--so the parts actually known of
the universe may comport many ideally possible complements.  But as
the facts are not the complements, so the knowledge of the one is
not the knowledge of the other in anything but the few necessary
elements of which all must partake in order to be together at all.
Why, if one act of knowledge could from one point take in the total
perspective, with all mere possibilities abolished, should there ever
have been anything more than that act?  Why duplicate it by the tedious unrolling, inch by inch, of the foredone reality?  No answer seems
possible. On the other hand, if we stipulate only a partial community of partially independent powers, we see perfectly why no one part controls the whole view, but each detail must come and be actually given, before, in any special sense, it can be said to be determined at all.  This is the moral view, the view that gives to other powers the same freedom it would have itself,---not the ridiculous `freedom to do right,' which in my mouth can only mean the freedom to do as {\it I\/} think right, but the freedom to do as {\it they\/} think right, or wrong either.
\eq

This is a world of ``objective indeterminism'' indeed, but one with no place for ``objective chance'' in the sense of Lewis's Principal Principle \cite{Lewis86a,Harper09}.  From within any part, the future is undetermined.\footnote{Indeed from the ``whole''---though in no real sense is there actually a whole---the future is undetermined as well, lest this vision fall back into a salacious monism.  Of course an Everettian will say, ``I have a whole that captures everything, leaving nothing unavailable: It is the universal wavefunction.''  But that is just the salacious monism we were talking about.  To make the point dramatic, we could quote William James's 1906 comment on the Hegelians of his own time \cite{JamesPragmatism}:
``The more absolutistic philosophers dwell on so high a level of
abstraction that they never even try to come down.  The absolute mind
which they offer us, the mind that makes our universe by thinking it,
might, for aught they show us to the contrary, have made any one of a
million other universes just as well as this.  You can deduce no
single actual particular from the notion of it.  It is compatible
with any state of things whatever being true here below.''  James could just as well have been talking about the Everett-style multiverse---as far as global philosophic goals go, there is little difference between these conceptions 103 years apart.  But the real point that is conclusive for us at this point of the text is simply that from any agent's perspective, there is uncertainty, full stop.}  If one of those parts is an agent, then it is an agent in a situation of uncertainty.  And where there is uncertainty, agents should use the calculus of Bayesian probability in order to make the best go at things.

But we have learned enough from Copernicus to know that egocentrism, whenever it can be shaken away from a {\it weltanschauung\/}, it ought to be.  Whenever ``I'' encounter a quantum system, and take an action upon it, it catalyzes a consequence in my experience that my experience could not have foreseen.  Similarly, by a Copernican principle, I should assume the same for ``you'':  Whenever you encounter a quantum system, taking an action upon it, it catalyzes a consequence in your experience.  By one category of thought, we are agents, but by another category of thought we are physical systems.  And when we take actions upon each other, the category distinctions are symmetrical.  Like with the Rubin vase, the best the eye can do is flit back and forth between the two formulations. In the common circles of the philosophy of science there is a strong popularity in the idea that agentialism can always be reduced to some complicated property arrived at from physicalism.  But perhaps this republican-banquet vision of the world that so seems to fit a QBist understanding of quantum mechanics is telling us that the appropriate ontology we should seek would treat these dual categories as just that, dual aspects of a higher, more neutral concept \cite{PureExperience}.  That is, these concepts ``action'' and ``unforseen consequence in experience,'' both so crucial for clarifying the very meaning of quantum measurement, might just be applicable after a fashion to arbitrary components of the world---i.e., venues in which probability talk has no place.   Understanding or rejecting this idea is the long road ahead of us.  The development of formalism in this paper we see as at least one promising entrance into that road.

We leave an old teacher of ours with the last word:\medskip
\begin{flushright}
\baselineskip=13pt
\parbox{2.8in}{\baselineskip=13pt\small
It is difficult to escape asking a challenging question. Is the
entirety of existence, rather than being built on particles or fields
of force or multidimensional geometry, built upon billions upon
billions of elementary quantum phenomena, those elementary acts of
``observer-participancy,'' those most ethereal of all the entities
that have been forced upon us by the progress of science?}
\medskip\\
\small --- John Archibald Wheeler, 1911--2008
\end{flushright}

\section{Acknowledgements}
We thank D. M. Appleby, G.~Bacciagaluppi, H.~C. von Baeyer, H.~Barnum, H.~R. Brown, W.~Demopoulos, S.~T. Flammia, L.~Hardy, P.~Hayden, K.~Martin, N.~D. Mermin, R.~Morris, W.~C. Myrvold, D.~Parker, R.~W. Spekkens, C.~Ududec, and J.~Uffink for numerous discussions.  We especially thank Robin Blume-Kohout for preventing us from egregious errors and, in consequence, inspiring Section \ref{Vivilify}, and {\AA}sa~Ericsson for many direct improvements to the evolving manuscript.  CAF dedicates the equations featured in Figure 2 to Geraldine Thigpen Spears on the occasion of her 80$^{\rm th}$ birthday. \medskip
 
This research was supported in part by the Perimeter Institute -- Australia Foundations Collaboration (PIAF) and in part by the U.~S. Office of Naval Research (Grant No.\ N00014-09-1-0247). Research at PI is supported by the Government of Canada through Industry Canada and by the Province of Ontario through the Ministry of Research \& Innovation. Research in Australia under the PIAF Collaboration is supported by Griffith U., U. of Queensland, U. of Sydney, and the Australian Research Council.

\end{document}